\documentclass[aps, prl, 10 pt, notitlepage, nofootinbib, twocolumn, letterpaper, superscriptaddress, preprintnumbers, longbibliography, floatfix]{revtex4-1}
\usepackage{epsfig,amsfonts,mathrsfs,amsmath,amssymb,graphicx,color,slashed,bbm}
\usepackage{amsmath,latexsym,amssymb,graphicx,slashed,hyperref,color,enumerate,url,etoolbox}

\definecolor{nicered}{rgb}{.7,.1,.1}
\definecolor{nicegreen}{rgb}{.1,.5,.1}
\definecolor{darkblue}{rgb}{0,0,.5}
\hypersetup{colorlinks, citecolor=nicegreen, linkcolor=nicered, urlcolor=darkblue}

\begin{document}

\title{Dark Matter Dilution Mechanism through the Lens of Large Scale Structure}

\author{Miha Nemev\v{s}ek}
\email{miha.nemevsek@ijs.si}
\affiliation{Jo\v{z}ef Stefan Institute, Jamova 39, 1000 Ljubljana, Slovenia}
\affiliation{Faculty of Mathematics and Physics, University of Ljubljana, Jadranska 19, 1000 Ljubljana, Slovenia}

\author{Yue Zhang}
\email{yzhang@physics.carleton.ca}
\affiliation{Department of Physics, Carleton University, Ottawa, ON K1S 5B6, Canada}

\date{\today}

\begin{abstract}

Entropy production is a necessary ingredient for addressing the over-population of thermal relics. It is widely employed in particle physics models for explaining the origin of dark matter. A long-lived particle that decays to the known particles, while dominating the universe, plays the role of the dilutor. We point out the impact of its partial decay to dark matter on the primordial matter power spectrum. For the first time, we derive a stringent limit on the branching ratio of the dilutor to dark matter from large scale structure observation using the SDSS data. This offers a novel tool for testing models with a dark matter dilution mechanism. We apply it to the left-right symmetric model and show that it firmly excludes a large portion of parameter space for right-handed neutrino warm dark matter.

\end{abstract}

\maketitle

%%%%%%%%%%%%%%%%%%%%%%%%%%%%%%%%%%%%%%%%%
\textbf{Introduction --}
%%%%%%%%%%%%%%%%%%%%%%%%%%%%%%%%%%%%%%%%%
The nature of dark matter (DM) is a tantalizing puzzle about our universe and new elementary particle(s) are arguably the 
most compelling DM candidates. Their non-gravitational interactions may establish thermal contact with the known particles 
in the early universe and dynamically produce the observed DM relic density.

The famous example of this kind is the weakly-interacting massive particle ``WIMP miracle'', where the DM mass is pinned around the electroweak scale by the freeze-out mechanism~\cite{Kolb:1990vq}. Going to smaller masses sometimes requires the existence of a new light dark force carrier, leading to the ``dark-sector'' theories~\cite{Alexander:2016aln}. Alternatively, if the mediator remains heavy, a light DM particle would freeze-out relativistically. If nothing else happens, such relic will either over-close the universe or remain too hot 
for structure formation. Moreover, excessive DM production could take place via feeble processes~\cite{Ellis:1982yb, Ellis:1983ew}. An attractive mechanism for addressing such over-abundance problems resorts to late time entropy production, injected by the decay of a long-lived particle~\cite{Scherrer:1984fd, Moroi:1999zb, Baltz:2001rq, Asaka:2006ek, Bezrukov:2009th, Nemevsek:2012cd, Arcadi:2011ev, Zhang:2015era, Patwardhan:2015kga, Soni:2017nlm, Contino:2018crt, Evans:2019jcs, Cosme:2020mck, Dror:2020jzy, Chanda:2021tzi, Asadi:2021bxp, Hasenkamp:2010if}, often referred to as the ``dilutor''. Such dilution mechanism has been utilized in a wide variety of theory contexts, e.g. sterile neutrino DM from gauge extensions~\cite{Bezrukov:2009th, Nemevsek:2012cd, Dror:2020jzy}, gravitino DM via freeze in/out production~\cite{Moroi:1999zb, Baltz:2001rq, Hasenkamp:2010if}, and glueball DM~\cite{Soni:2017nlm}.

In this letter, we point out and investigate an important aspect of generic DM dilution mechanisms -- the (subdominant) decay of the dilutor into DM. Such channels hamper the goal of diluting the DM relic density and are often omitted in simplified analyses. However, they do arise in well-motivated ultraviolet complete theories, either at tree level or via radiative corrections, with its branching ratio fixed by the theory structure. This ratio serves as a key parameter that characterizes the dilution mechanism.

Our main result is to show that DM produced by the dilutor decay is typically energetic and can have a profound impact on the primordial matter power spectrum. We show that the existing large scale structure (LSS) data already constrains the branching ratio of this decay mode to percent level. This is a general result and serves as a goalpost for various DM dilution mechanisms. A concrete example of right-handed neutrino DM from the minimal Left-Right symmetric model (LRSM)~\cite{Mohapatra:1974hk, Senjanovic:1978ev, Mohapatra:1979ia}, will be discussed.

\begin{figure}[htp]
  \begin{center}
    \includegraphics[width=0.45\textwidth]{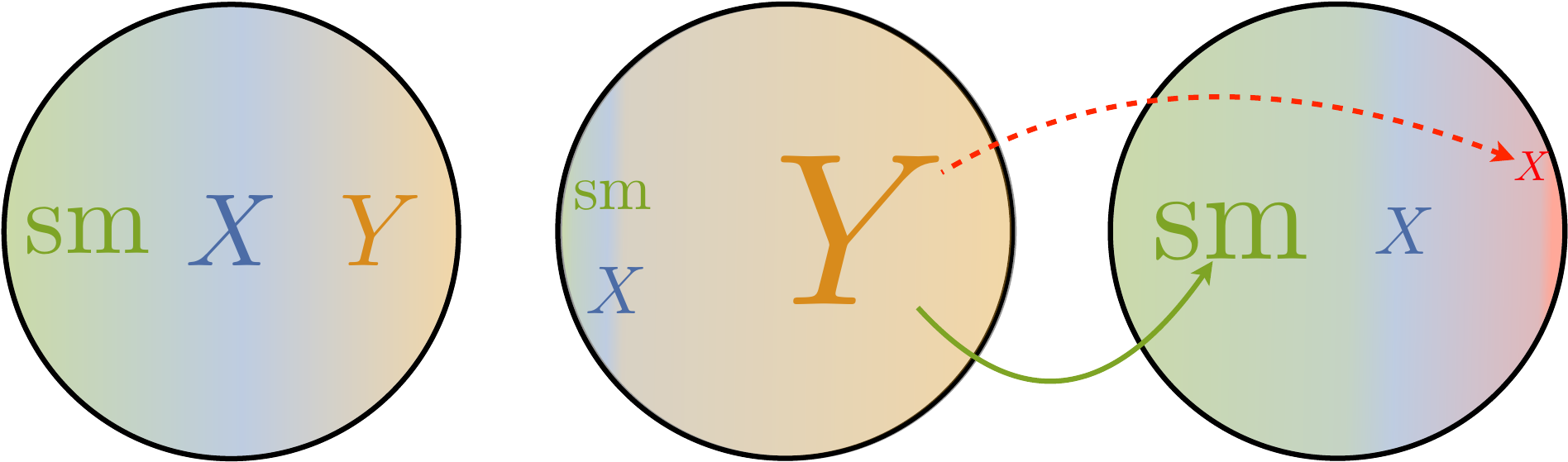}
  \end{center}
  \caption{Schematic illustration of the DM dilution mechanism: relativistic freeze-out, dilutor matter
  domination, entropy injection (dashed red) and DM re-population (solid green).}
\label{fig:Illust}
\end{figure}

%%%%%%%%%%%%%%%%%%%%%%%%%%%%%%%%%%%%%%%%%%%%%%
\textbf{Entropy Dilution for Relic Density --}
%%%%%%%%%%%%%%%%%%%%%%%%%%%%%%%%%%%%%%%%%%%%%%
Hereafter, we refer to the DM as $X$ and the dilutor as $Y$. Both are in thermal equilibrium with the visible sector at early times and decouple relativistically
at similar temperatures. This makes their initial abundances equal, barring spin factors. The correct DM relic density is obtained if the dilutor $Y$ is sufficiently long-lived, 
becomes non-relativistic and dominates the energy density before decaying away, as shown in FIG.~\ref{fig:Illust}.

The dilutor $Y$ reheats the universe by decaying into the visible final states ``SM'', however it can also decay into $X$,
\begin{align}\label{eq:decaychannels}
    Y &\to SM \ , 
    &
    Y &\to n X \ (\ +\ m \ SM \ ) \ ,
\end{align}
where $n, m \in \mathbb Z^+$ count the multiplicity of $X$ and SM particles, respectively. The first decay channel dumps entropy into the visible sector and dilutes the primordial 
thermal population of $X$. The second channel, whose branching ratio is given by ${\rm Br}_X$, produces a secondary 
non-thermal population of $X$ that repopulates DM and contributes to its final relic density. The bracket in~\eqref{eq:decaychannels} includes the possibility that the secondary channel 
is completely dark, without any ``SM'' in the final state. In the absence of extended dark sectors, the branching ratio of the first channel is $1 - {\rm Br}_X$.

After dilution, DM becomes non-relativistic with the relic abundance~\cite{Bezrukov:2009th, Nemevsek:2012cd}
\begin{equation}\label{eq:DMrelic}
  \Omega_X \simeq 0.26 \left( 1 + n {\rm Br}_X \right) 
  \left( \frac{m_X}{1\,\text{keV}} \right) \left( \frac{1\,\text{GeV}}{m_Y} \right)
  \sqrt{\frac{1\,\text{sec}}{\tau_Y}} \ ,
\end{equation}
derived under the sudden decay approximation~\cite{Scherrer:1984fd}. Using $H(T_{\rm RH}) \tau_Y = 1$, $Y$ 
reheats the universe to a temperature
\begin{equation}\label{eq:TRH}
  T_{\rm RH} \simeq \frac{1\,{\rm MeV}}{g_*(T_{\rm RH})^{1/4}} 
  \left( \frac{m_Y}{10^6 m_X} \right) \ ,
\end{equation}
where the lifetime of $Y$ is determined by $m_Y/m_X$ from~\eqref{eq:DMrelic}.
For successful big-bang nucleosynthesis (BBN), we need $T_{\rm RH} > \text{MeV}$,
which requires a large hierarchy between the dilutor and DM masses $m_{Y} \gg m_{X}$.

Such scenarios feature two distinct populations of $X$: the primordial and the secondary one. The thermal primordial component gets cooled by entropy production from $Y$ decays
below the photon temperature. Conversely, the secondary $X$ are more energetic and carry the energy on the order of $m_Y/(n+m) \gg T_X$, where $T_X$ is the temperature of the primordial $X$. We assume that $X$ remains collision-less after $Y$ decay.

Consider the phase space distribution of $X$ particles, $f_X(x, t)$, which is a function of time $t$ and $x = E_X/T_X$. With this convention, the $f_X(x, t)$ satisfies
\begin{equation}
  \frac{T_X^3}{2\pi^2} x^2 \dot{f}_X(x,t) = \frac{\rho_Y(t) \Gamma_Y 
  {\rm Br}_X}{m_Y}\frac{T_X}{m_Y} g\left( \omega \right) \, ,
\end{equation}
where $\omega = E_X/m_Y$. The $g(\omega)$ function is the distribution of the energy fraction carried by $X$ in the rest frame of $Y$. Its form is model dependent and normalized to $\int {\rm d} \omega g(\omega) = n$. Averaging the energy over $g(\omega)$ gives
\begin{align}
  y = \int {\rm d}\omega \, \omega  \, g(\omega) \ , 
\end{align}
which characterizes the energy fraction carried by $X$ in the $Y \to n X (+ m\,SM)$ decay.

For concreteness we consider the two models:
\begin{enumerate}
  \item The LRSM, where $X$ is the lightest right-handed neutrino $N_1$ and $Y$ is a heavier $N_2$, which undergoes a three-body decay into $N_1$ plus two charged leptons, mediated by the $W_R$ gauge boson~\cite{Bezrukov:2009th, Nemevsek:2012cd, Dror:2020jzy}; 
  \item A long-lived scalar $\Phi$ with a partial decay width into two fermionic DM. $\Phi$ may be incarnated as a modulus field in supersymmetric theories~\cite{Moroi:1999zb, Acharya:2009zt}. 
\end{enumerate}
The corresponding $g$ functions and $n, y$ integrals are summarized in TAB.~\ref{tab:models}. The masses of final-state charged leptons are neglected here.

\begin{table}[h]
  \centering
  \begin{tabular}{|c|c|c|c|c|}
  \hline
                    & $n$   & $g(\omega)$ & $\,\,\omega_{\rm max}\,\,$ & $y$ \\
  \hline
  LRSM              & 1     & $\,\,16 \omega^2 (3-4\omega)\,\,$ & 1/2 & 7/20 \\
  \hline
  Long-lived scalar & 2     & $2 \delta (\omega - 1/2)$ & $-$ & 1 \\
  \hline
  \end{tabular}
  \caption{The energy fraction distribution $g(\omega)$, taken away by $X$ in the rest frame of $Y$, and its integrals $n = \int g, y = \int \omega g$. The first row applies to the LRSM, where a single $X$ ($n=1$) is produced via a 3-body decay.
  The second corresponds to a long-lived scalar, where $\Phi \to X X$ produces two DM states ($n=2$), with $\delta$ being the Dirac delta function.}
 \label{tab:models}
\end{table}

\begin{figure}[t]
  \begin{center}
  \vspace{-0.4cm}
    \includegraphics[width=0.5\textwidth]{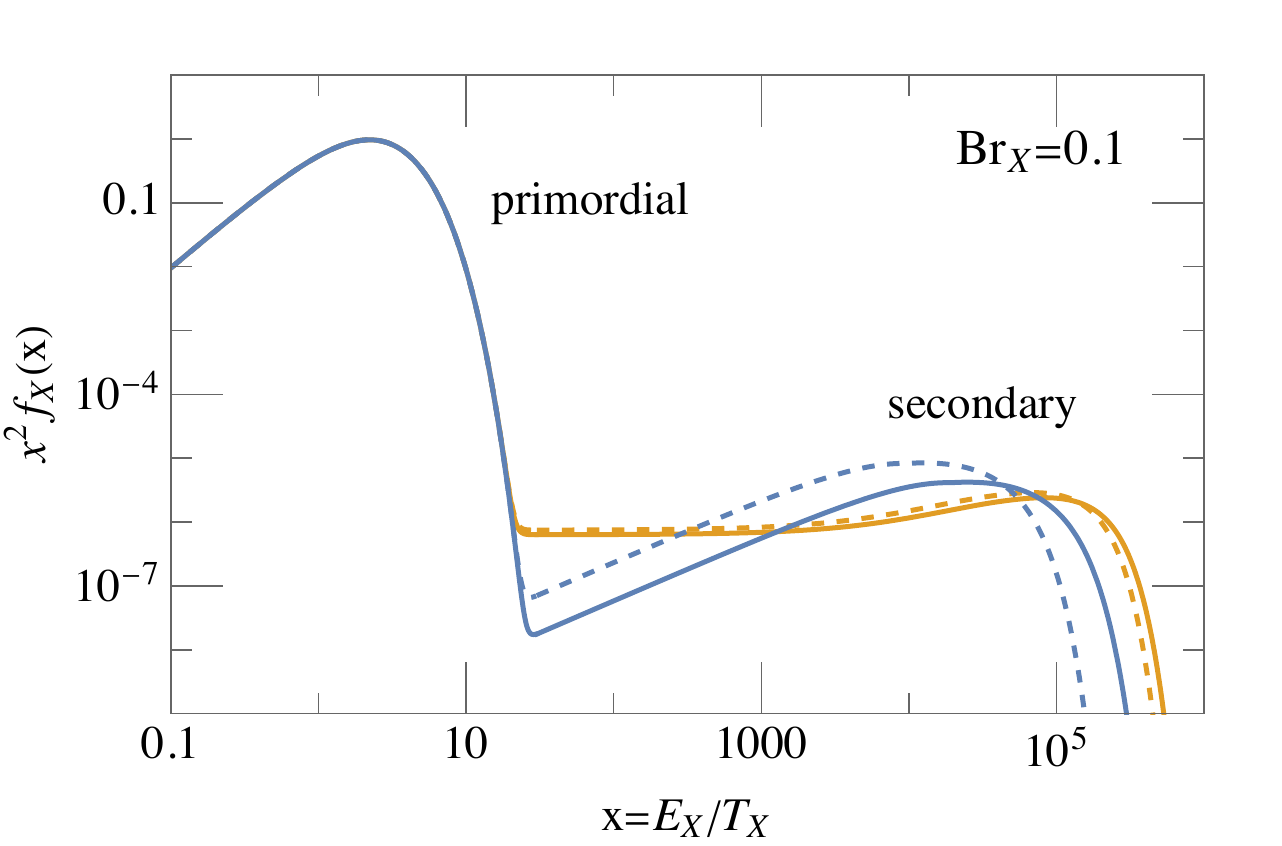}
  \end{center}
  \caption{Phase space distribution of ultra-relativistic $X$. The blue and orange curves correspond to the two models listed in TAB.~\ref{tab:models}. Solid (dashed) curve corresponds to $m_X = 10 \,{\rm keV}$, $m_Y = 100\,{\rm GeV}$ ($m_Y = 1 \,{\rm PeV}$), while ${\rm Br}_X = 0.1$ for all cases.}
\label{fig:fX}
\end{figure}

The time-dependence in the energy density $\rho_Y$ can be computed by solving the coupled 
Boltzmann equations
\begin{align}
  \dot \rho_Y + 3 H \rho_Y &= - \Gamma_Y \rho_Y \ , \\
  \dot \rho_X + 4 H \rho_X &= y {\rm Br}_X \Gamma_Y \rho_Y \ , \\
  \dot \rho_{\rm SM} + 4 H \rho_{\rm SM} &\simeq (1-y {\rm Br}_X) \Gamma_Y \rho_Y \ , \label{eq:8}
\end{align}
where $\rho_{\rm SM}$ is the energy density carried by relativistic visible particles and $H^2 = 8\pi G_N (\rho_Y + \rho_X+\rho_{\rm SM})/3$ is the Hubble parameter. This set of equations applies for non-relativistic $Y$, while the $X$ population remains ultra-relativistic. We neglect the temperature dependence of $g_*$ in Eq.~\eqref{eq:8}.
Defining $\zeta = \ln a$ to keep track of time, the $f_X$ can be solved as
\begin{equation} \label{eq:fX}
\begin{split}
  f_X(x) &= \frac{1}{e^x + 1}
  \\
  &+ \frac{2\pi^2}{x^2} \frac{\Gamma_Y {\rm Br}_X}{m_Y^2} 
  \int {\rm d} \zeta \, \frac{\rho_Y(\zeta)}{T_X^2 H(\zeta)} g \left(\frac{T_X}{m_Y} x \right) \ ,
\end{split}
\end{equation}
as illustrated on FIG.~\ref{fig:fX}. The first term of~\eqref{eq:fX} is the primordial Fermi-Dirac distribution of $X$ and the second is the non-thermal re-population of $X$, where the $\zeta$ integral goes from $T=m_Y/10$ to $t = 10 \, \tau_Y$ that covers the entire period of $Y$ decay. 

The secondary component of $X$ has a significantly smaller occupancy, but carries more energy and thereby affects structure formations. The shape of the secondary component is roughly independent of $m_Y$. This follows from Eqs.~\eqref{eq:DMrelic} and~\eqref{eq:TRH}, where the relic density requires the scaling $m_Y \sim 1/\sqrt{\tau_Y}$ and the reheating temperature is set by the Hubble time $T_{\rm RH} \sim \sqrt{H} \sim 1/\sqrt{\tau_Y}$. Immediately after $Y\to X$ decay, $E_X \lesssim m_Y$ and $T_X \sim T_{\rm RH}$. As a result, the kinematic endpoint $x_{\rm max}\sim m_Y/T_{\rm RH}$ is held constant for fixed $m_X$, irrespective of $m_Y$ or $\tau_Y$.

\begin{figure}[t]
  \begin{center}
    \includegraphics[width=0.47\textwidth]{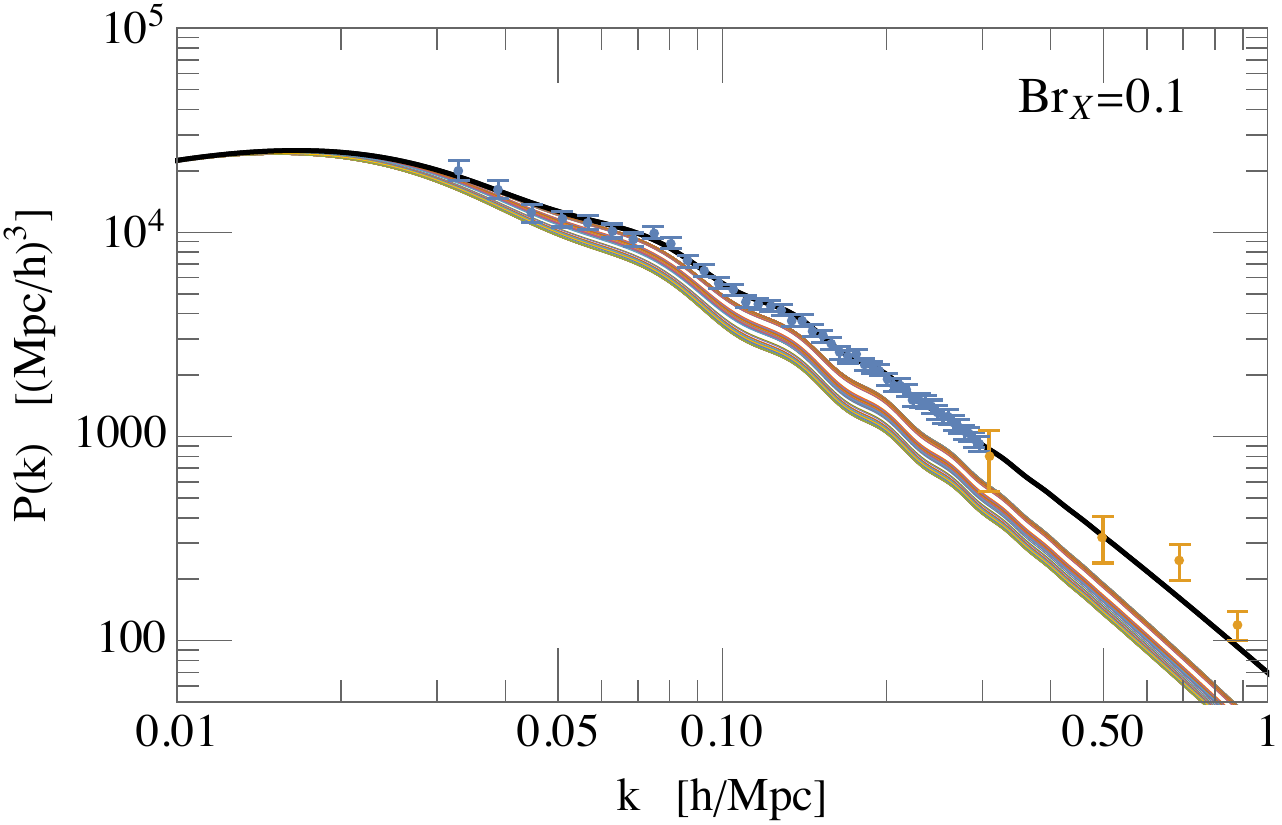}
  \end{center}
  \caption{Primordial matter power spectrum in standard $\Lambda$CDM (black solid curve) and diluted DM models listed in TAB.~\ref{tab:models} (colorful curves). Like FIG.~\ref{fig:fX}, we set ${\rm Br}_X = 0.1$. Data points from SDSS DR7 LRG and Lyman-$\alpha$ observations are shown in blue and orange, respectively.}
\label{fig:pk}
\end{figure}

%%%%%%%%%%%%%%%%%%%%%%%%%%%%%%%%%%%%%%%%%%%%%%%%
\textbf{Imprint on the Matter Power Spectrum --}
%%%%%%%%%%%%%%%%%%%%%%%%%%%%%%%%%%%%%%%%%%%%%%%%
Let us begin with a simple heuristic understanding of how the matter power spectrum is affected, which we then sharpen by a thorough numerical study. The secondary population of $X$ becomes non-relativistic when the photon temperature drops to
\begin{equation}\label{eq:TNR}
  T_{\rm NR} \sim  1\,{\rm eV} \, (n+m) \, g_{*}(T_{\rm RH})^{1/12} \ ,
\end{equation}
which is rather low for $n+m \sim \mathcal{O}(1)$ and nearly independent of $g_*$. Above $T_{\rm NR}$, the DM fluid is made of the non-relativistic primordial and the relativistic secondary component. The energy density of the latter is more important at temperatures above $T_{\rm NR}/{\rm Br}_X$. In this regime, the effective sound speed of the $X$ fluid is large, which interrupts the 
regular logarithmic growth of density perturbations in $X$. This suppresses the primordial matter power spectrum $P(k)$ on length scales smaller than the Hubble radius at temperature $T_{\rm NR}/{\rm Br}_X$, see FIG.~\ref{fig:pk}. The resulting $P(k)$ would thus disagree with the LSS measurements, unless ${\rm Br}_X \ll 1$.

We now turn to a quantitative numerical analysis in the parameter space of $m_X$ versus $m_Y$. For each point we first set the dilutor lifetime $\tau_Y$ using Eq.~\eqref{eq:DMrelic}. Next, we determine the phase space distribution $f_X$ with Eq.~\eqref{eq:fX} and evolve it with {\tt CLASS}~\cite{Lesgourgues:2011re, Blas:2011rf, Lesgourgues:2011rh} to obtain $P(k)$. We scan over $200$ points in the mass range $m_X \in (1\,{\rm keV}, 1\,{\rm MeV})$ and $m_Y \in (1\,{\rm GeV}, 10^{16}\,{\rm GeV})$ for both models in TAB.~\ref{tab:models}. The results are shown by the colored curves in FIG.~\ref{fig:pk}, where we set ${\rm Br}_X = 0.1$, while the black solid curve is the fiducial $\Lambda$CDM model.

\begin{figure*}
  \begin{center}
    \includegraphics[width=0.45\textwidth]{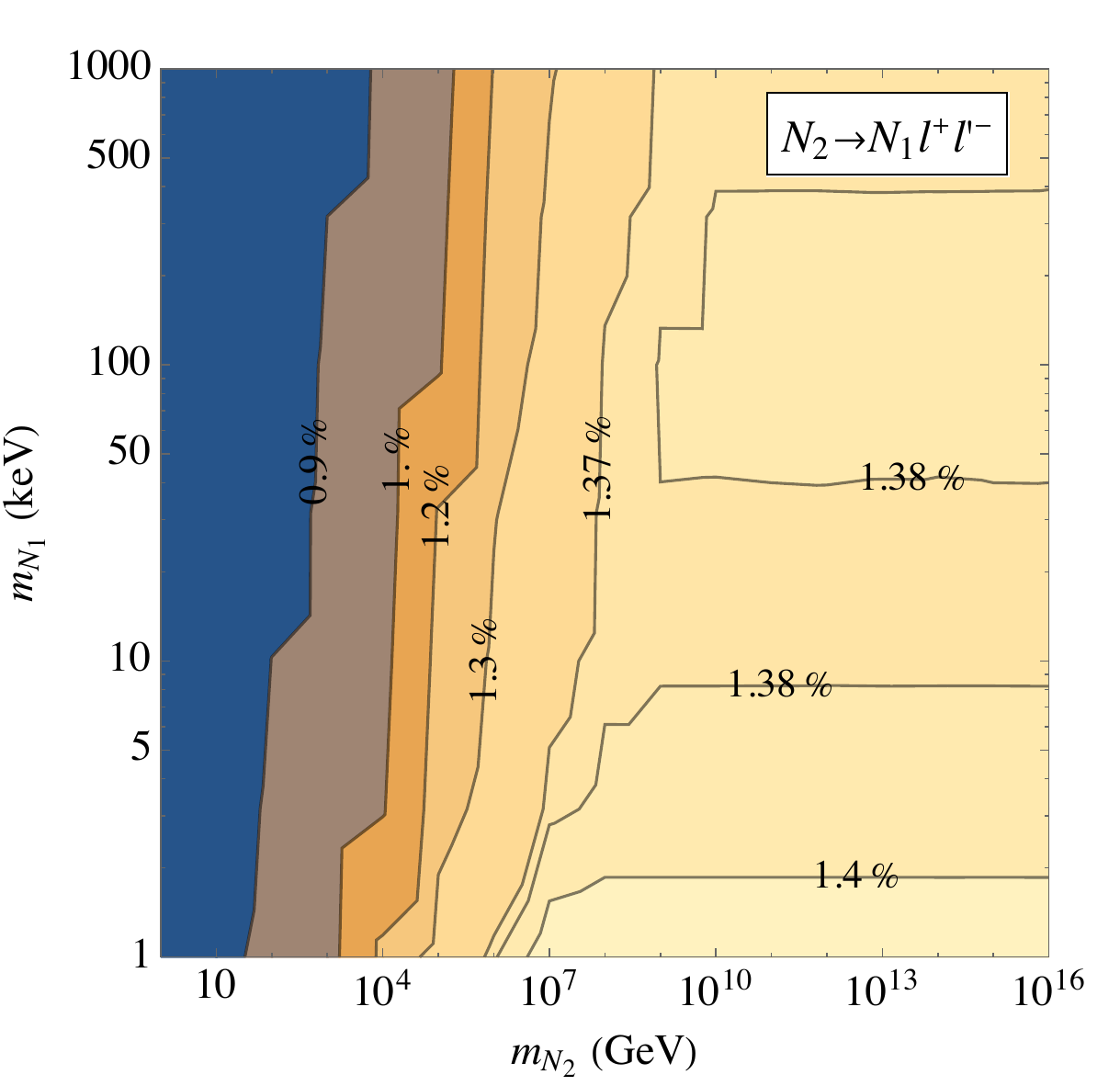} \quad
    \includegraphics[width=0.45\textwidth]{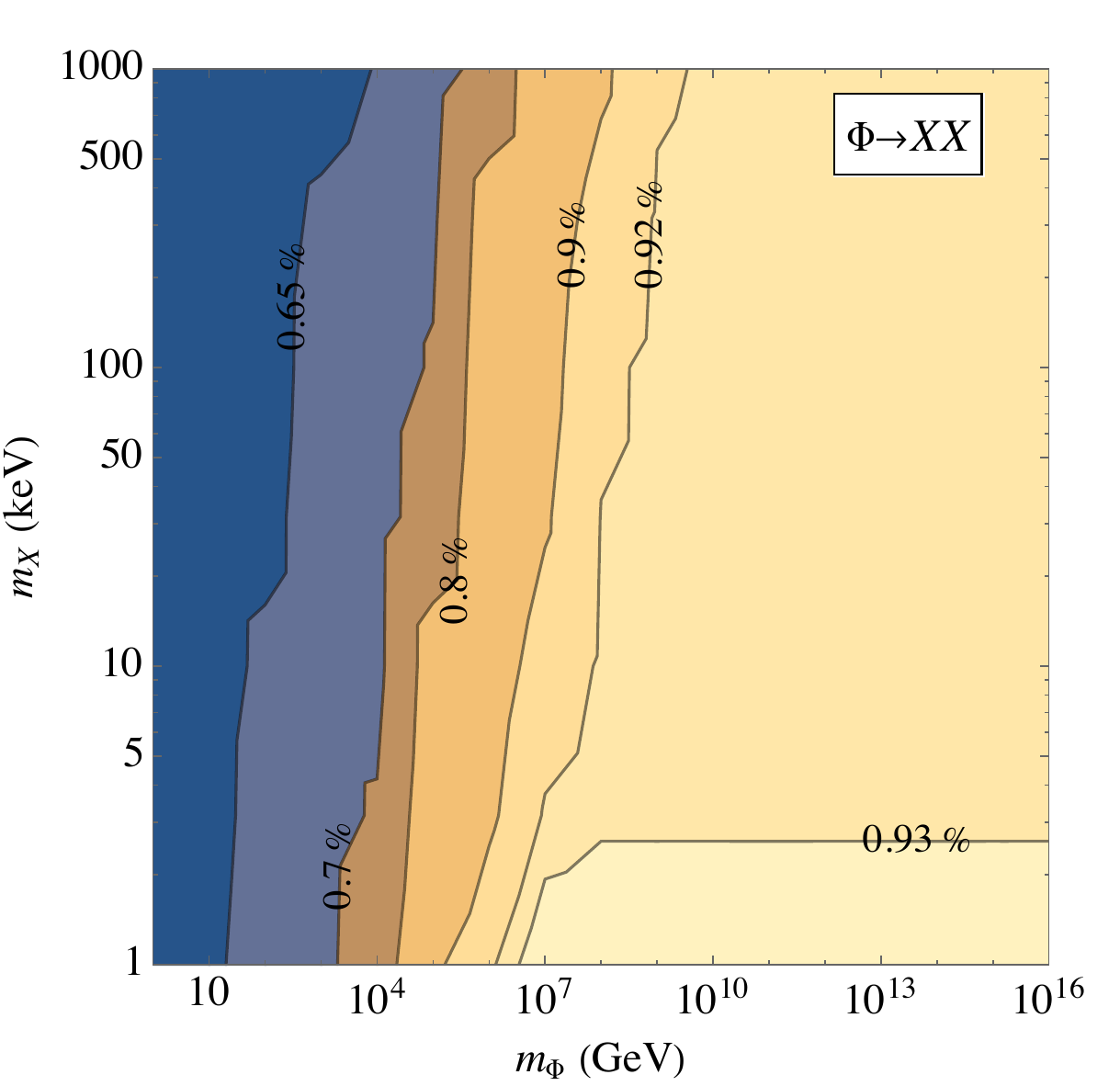}
  \end{center}
  \caption{
  Upper bound on $\text{Br}_X$, the branching ratio of dilutor $Y$ decaying into $X$ from the fit to LSS data (SDSS DR7 LRG), for the LRSM with $Y = N_2$, $X = N_1$, and a long-lived scalar $Y = \Phi$. For each point in the $m_X - m_Y$ parameter space, the $Y$ lifetime is solved by requiring $X$ to comprise all of the DM in the universe.}
\label{fig:maxBr}
\end{figure*}

The experimental data points come from the SDSS DR7 on luminous red galaxies~\cite{Reid:2009xm} (blue) and the Lyman-$\alpha$ forest~\cite{Lee:2012xb} (orange) measurements. All the curves in scenarios with secondary $X$ share a common feature with significant deviations from data in the $k \gtrsim 0.03\, h/{\rm Mpc}$ region. These occur at a much lower $k$ compared to other DM production mechanisms such as thermal freeze-in~\cite{DEramo:2020gpr, Decant:2021mhj}. This is mainly due to the large hierarchy between the dilutor and DM mass, required by Eqs.~\eqref{eq:DMrelic} and~\eqref{eq:TRH}. Based on a simple $\Delta \chi^2$ fit to the data, we find that the LSS data from SDSS sets a much stronger constraint on these scenarios than Lyman-$\alpha$, making this probe particularly robust. The conflict with data increases with ${\rm Br}_X$, which translates into an upper bound in FIG.~\ref{fig:maxBr} for the two models in TAB.~\ref{tab:models}. The result is similar for both cases and the branching ratio of the dilutor decaying into DM is constrained to be 
\begin{equation}\label{eq:maxBr}
  {\rm Br}_X \lesssim 1 \% \ , \quad @ 95\% \text{ CL} \ .
\end{equation}

This bound is nearly independent of $m_Y$. It gets slightly relaxed for larger $m_Y$, because holding the DM relic density fixed in Eq.~\eqref{eq:DMrelic} requires the $\tau_Y$ to be shorter, leading to a higher reheating temperature after the decay of dilutor. The corresponding temperature for the secondary DM component to become non-relativistic also increases, which is a $\sqrt[12]{g_*}$ effect \`a la Eq.~\eqref{eq:TNR}. Eventually, this shifts the deviation of $P(k)$ to a slightly higher $k$ and becomes less constraining.

The constraint derived here comes predominantly from the LSS data, which relies only on the evolution of matter density perturbations in the linear regime. LSS thus provides a robust test of these models and we expect similar constraints to apply broadly for models that utilize the dilution mechanism for addressing the DM relic density.

Our result can be generalized to initial abundances for $Y$ and $X$ beyond the relativistic freeze-out. A sub-thermal initial population of $Y$ needs to be heavier and/or longer lived in order to 
provide the same amount of entropy injection. The secondary $X$ particles from $Y$ decay are more energetic and take longer to become matter-like. This impacts the primordial matter power spectrum down to even lower $k$ and leads to a more stringent constraint on ${\rm Br}_X$ than Eq.~\eqref{eq:maxBr}. On the contrary, starting with a smaller overpopulation of $X$, the constraint on ${\rm Br}_X$ 
will be weaker.

%%%%%%%%%%%%%%%%%%%%%%%%%%%%%%%%%%%%%%%%%%%%%%
\textbf{Implications for Left-Right Symmetry --}
%%%%%%%%%%%%%%%%%%%%%%%%%%%%%%%%%%%%%%%%%%%%%%
The LRSM is an elegant framework to accommodate the right-handed neutrino $N$ as DM. Gauging the $SU(2)_L\times SU(2)_R \times U(1)_{B-L}$ necessitates three
$N$s for anomaly cancellation. The lightest $N_1$, if cosmologically long-lived, is a DM candidate. Assuming the early universe was sufficiently hot, $N_1$ would be kept in thermal equilibrium with 
the known particles by the $SU(2)_R$ gauge interactions and would freeze-out relativistically. Late entropy production is then required to set the relic density of $N_1$ to the observed DM value.

One (or both) of the heavier right-handed neutrinos, $N_2$, can play the role of the dilutor~\cite{Bezrukov:2009th}. In the simplest scenario, $N_2$ decays through the exchange of $W_R$, 
in close analogy to heavy quarks in the Standard Model,
\begin{align}\label{eq:LRN2decay}
    N_2 &\to \ell q \bar q' \ , 
    &
    N_2 &\to N_1 \ell \ell' \ ,
\end{align}
where we assumed that $m_{N_2} > m_\tau$, such that all leptonic channels are open. The first channel plays the desired role of dilution, whereas the second produces energetic $N_1$ particles that affect the matter power spectrum, as discussed earlier. For $N_2$ mass well above the electroweak scale, the branching ratio for $N_2 \to N_1 \ell \ell'$ is 1/10. If the mass of $N_2$ is below the top quark mass, the ratio grows to 1/7. Such values are in complete contradiction with the main findings of this work in Eq.~\eqref{eq:maxBr} and FIG.~\ref{fig:maxBr}. In this scenario, the entropy dilution mechanism for addressing the DM relic density is thus firmly excluded.

Such a strong constraint can only be evaded with additional non-DM decays of $Y$. The $N_2$ could decay via the SM $W$, mediated either by $N_2$ and light neutrino $\nu$ mixing, or the $W$-$W_R$ gauge boson mixing. Neutrino mixing comes about if $N_2$ participates in the type-I seesaw mechanism~\cite{Mohapatra:1980yp} and is uniquely fixed by $\theta_{N_2\nu} \simeq \sqrt{m_\nu/m_{N_2}}$~\cite{Nemevsek:2012iq}. 
In the presence of the $N_2$-$\nu$ mixing alone, we find that $N_2$ must be lighter than the $W$ boson for it to be a long-lived dilutor~\cite{wip2022}. For the $W$ mediated decay channels to dominate over the $W_R$ mediated ones in order to satisfy the LSS constraint derived in Eq.~\eqref{eq:maxBr}, we need
\begin{equation}
  M_{W_R}\gtrsim 55 \text{ TeV} \left( \frac{m_{N_2}}{1 \text{ GeV}} \right)^{1/4} \ ,
\end{equation}
where we set $m_\nu = \sqrt{\Delta m^2_{\rm at}}\simeq 0.05$~eV. 
On the other hand, the $W$-$W_R$ mixing is of order $\xi_{\rm LR} \sim M_W^2/M_{W_R}^2$. In this case, $N_2$ must be heavier than the $W$ boson in order to satisfy the LSS constraint. Producing the correct DM relic density then requires $M_{W_R} > 100\,$TeV~\cite{wip2022}.

It is remarkable that in all cases cosmological measurement can set a much stronger limit on the LRSM than the high-energy colliders~\cite{Keung:1983uu, Nemevsek:2011hz, Nemevsek:2018bbt} with the latest limits~\cite{CMS:2016ifc, ATLAS:2019lsy, ATLAS:2019isd, CMS:2021dzb, Abdullahi:2022jlv} at $M_{W_R} > 5.6 \text{ TeV}$, and supernova cooling~\cite{Barbieri:1988av}. Another way of suppressing the re-population of DM $N_1$ concerns the flavor structure of the 
right-handed lepton mixing matrix and the mass spectrum of $N_i$~\cite{Nemevsek:2012cd, wip2022}.

%%%%%%%%%%%%%%%%%%%%%%%%%%%%%%%%%%%%%%%%%%%%%%%%%
\textbf{The $\Delta N_{\text{eff}}$ prediction --}
%%%%%%%%%%%%%%%%%%%%%%%%%%%%%%%%%%%%%%%%%%%%%%%%%
The secondary $X$ particles from $Y$ decay contribute to the total energy density of relativistic fluids in the universe until $T_{\rm NR} \sim {\rm eV}$. This could lead to a deviation of $\Delta N_{\rm eff}$ from the standard $\Lambda$CDM model. At early times, such as the BBN era, the contribution to $\Delta N_{\rm eff}$ can be derived in terms of parameters of the dilution mechanism
\begin{equation}
  \Delta N_{\rm eff} \simeq \frac{43}{7} \frac{y \, {\rm Br}_X}{1 - y \, {\rm Br}_X} 
  \left( \frac{43}{4 \, g_{*}(T_{\rm RH})} \right)^{1/3} \ ,
\end{equation}
where the $y {\rm Br}_X/(1 - y {\rm Br}_X)$ factor corresponds to the energy density ratio of the $X$ fluid and the visible sector, immediately after $Y$ decay. Assuming the constraint in Eq.~\eqref{eq:maxBr} is saturated, we find $\Delta N_{\rm eff} \lesssim 0.062$. This is too small to affect the BBN, but might become relevant for precision cosmological measurements at the future CMB$-$Stage~4~\cite{CMB-S4:2016ple}.

%%%%%%%%%%%%%%%%%%%%%%%%%%%%%%%%%%%%%%%%%%%%%%
\textbf{On the Lower Bound for the DM Mass --}
%%%%%%%%%%%%%%%%%%%%%%%%%%%%%%%%%%%%%%%%%%%%%%
Let us comment on the limit when ${\rm Br}_X \to 0$, i.e., no DM is produced in the dilutor decay. In this case, all DM in the universe is made of the primordial $X$, which follows a thermal
distribution, but has a lower temperature than photons due to entropy production via $Y$ decay. It depends solely on the DM mass, 
\begin{equation}
  T_X / T_\gamma = 0.16\, \left( \frac{1\,{\rm keV}}{m_X} \right)^{1/3} \ .
\end{equation} 
Because of the lower temperature ratio compared to alternative production mechanisms, such as the oscillation production for sterile neutrino DM~\cite{Dodelson:1993je, DeGouvea:2019wpf, Kelly:2020pcy, Kelly:2020aks, Chichiri:2021wvw, Benso:2021hhh}, the regular warm DM constraints are weaker~\cite{Asaka:2006ek, Bezrukov:2009th, Nemevsek:2012cd, Patwardhan:2015kga, Evans:2019jcs, Dror:2020jzy}. Lyman-$\alpha$~\cite{Irsic:2017ixq}, Milky Way satellite galaxy~\cite{Nadler:2019zrb, DES:2020fxi}, strong lensing observations~\cite{Enzi:2020ieg} and phase space of dwarf galaxies~\cite{TremaineGunn, Boyarsky:2008ju, Domcke:2014kla, DiPaolo:2017geq} set a lower bound on DM mass of several keV.

%%%%%%%%%%%%%%%%%%%%%%%%%%%%%%%%%%%%%%%%%%%
\textbf{Outlook --}
%%%%%%%%%%%%%%%%%%%%%%%%%%%%%%%%%%%%%%%%%%%
We explore entropy dilution mechanisms for DM relic density by pointing out and quantifying the signature of dilutor decay to DM on the formation of LSS of the universe. Such a decay mode is common in DM models and the existing cosmological data can set a stringent constraint on its branching ratio. Our result mostly relies on LSS data, which belongs to the linear regime of structure growth and is thus very robust. It can be further improved by future LSS surveys~\cite{Ferraro:2022cmj, Chen:2016vvw}. It is also highly complementary to probes of structure formation on small scales and the cosmic microwave background. Our finding offers a new tool for testing and distinguishing particle physics models for diluted DM.

\begin{acknowledgements}
We thank Jeff Dror, Miguel Escudero and Nicholas Rodd for helpful discussions. MN is supported by the Slovenian Research Agency under the research core funding No. P1-0035 and in part by the research grants J1-3013, N1-0253 and J1-4389. Y.Z. is supported by the Arthur B. McDonald Canadian Astroparticle Physics Research Institute, and the Munich Institute for Astro, -Particle and BioPhysics (MIAPbP), which is funded by the Deutsche Forschungsgemeinschaft (DFG, German Research Foundation) under Germany's Excellence Strategy – EXC-2094 – 390783311. Y.Z. is grateful to the organisers of the Pollica Summer Workshop supported by the Regione Campania, Università degli Studi di Salerno, Università degli Studi di Napoli “Federico II”, i dipartimenti di Fisica “Ettore Pancini”  and “E R Caianiello”, and Istituto Nazionale di Fisica Nucleare.
\end{acknowledgements}

\bibliography{References}

\begin{thebibliography}{61}%
\makeatletter
\providecommand \@ifxundefined [1]{%
 \@ifx{#1\undefined}
}%
\providecommand \@ifnum [1]{%
 \ifnum #1\expandafter \@firstoftwo
 \else \expandafter \@secondoftwo
 \fi
}%
\providecommand \@ifx [1]{%
 \ifx #1\expandafter \@firstoftwo
 \else \expandafter \@secondoftwo
 \fi
}%
\providecommand \natexlab [1]{#1}%
\providecommand \enquote  [1]{``#1''}%
\providecommand \bibnamefont  [1]{#1}%
\providecommand \bibfnamefont [1]{#1}%
\providecommand \citenamefont [1]{#1}%
\providecommand \href@noop [0]{\@secondoftwo}%
\providecommand \href [0]{\begingroup \@sanitize@url \@href}%
\providecommand \@href[1]{\@@startlink{#1}\@@href}%
\providecommand \@@href[1]{\endgroup#1\@@endlink}%
\providecommand \@sanitize@url [0]{\catcode `\\12\catcode `\$12\catcode
  `\&12\catcode `\#12\catcode `\^12\catcode `\_12\catcode `\%12\relax}%
\providecommand \@@startlink[1]{}%
\providecommand \@@endlink[0]{}%
\providecommand \url  [0]{\begingroup\@sanitize@url \@url }%
\providecommand \@url [1]{\endgroup\@href {#1}{\urlprefix }}%
\providecommand \urlprefix  [0]{URL }%
\providecommand \Eprint [0]{\href }%
\providecommand \doibase [0]{http://dx.doi.org/}%
\providecommand \selectlanguage [0]{\@gobble}%
\providecommand \bibinfo  [0]{\@secondoftwo}%
\providecommand \bibfield  [0]{\@secondoftwo}%
\providecommand \translation [1]{[#1]}%
\providecommand \BibitemOpen [0]{}%
\providecommand \bibitemStop [0]{}%
\providecommand \bibitemNoStop [0]{.\EOS\space}%
\providecommand \EOS [0]{\spacefactor3000\relax}%
\providecommand \BibitemShut  [1]{\csname bibitem#1\endcsname}%
\let\auto@bib@innerbib\@empty
%</preamble>
\bibitem [{\citenamefont {Kolb}\ and\ \citenamefont
  {Turner}(1990)}]{Kolb:1990vq}%
  \BibitemOpen
  \bibfield  {author} {\bibinfo {author} {\bibfnamefont {E.~W.}\ \bibnamefont
  {Kolb}}\ and\ \bibinfo {author} {\bibfnamefont {M.~S.}\ \bibnamefont
  {Turner}},\ }\href {\doibase 10.1201/9780429492860} {\emph {\bibinfo {title}
  {{The Early Universe}}}},\ Vol.~\bibinfo {volume} {69}\ (\bibinfo
  {publisher} {{CRC Press}},\ \bibinfo {year} {1990})\BibitemShut {NoStop}%
\bibitem [{\citenamefont {Alexander}\ \emph {et~al.}(2016)\citenamefont
  {Alexander} \emph {et~al.}}]{Alexander:2016aln}%
  \BibitemOpen
  \bibfield  {author} {\bibinfo {author} {\bibfnamefont {J.}~\bibnamefont
  {Alexander}} \emph {et~al.}\ }(\bibinfo {year} {2016})\ \Eprint
  {http://arxiv.org/abs/1608.08632} {arXiv:1608.08632 [hep-ph]} \BibitemShut
  {NoStop}%
\bibitem [{\citenamefont {Ellis}\ \emph {et~al.}(1982)\citenamefont {Ellis},
  \citenamefont {Linde},\ and\ \citenamefont {Nanopoulos}}]{Ellis:1982yb}%
  \BibitemOpen
  \bibfield  {author} {\bibinfo {author} {\bibfnamefont {J.~R.}\ \bibnamefont
  {Ellis}}, \bibinfo {author} {\bibfnamefont {A.~D.}\ \bibnamefont {Linde}}, \
  and\ \bibinfo {author} {\bibfnamefont {D.~V.}\ \bibnamefont {Nanopoulos}},\
  }\href {\doibase 10.1016/0370-2693(82)90601-3} {\bibfield  {journal}
  {\bibinfo  {journal} {Phys. Lett. B}\ }\textbf {\bibinfo {volume} {118}},\
  \bibinfo {pages} {59} (\bibinfo {year} {1982})}\BibitemShut {NoStop}%
\bibitem [{\citenamefont {Ellis}\ \emph {et~al.}(1984)\citenamefont {Ellis},
  \citenamefont {Hagelin}, \citenamefont {Nanopoulos}, \citenamefont {Olive},\
  and\ \citenamefont {Srednicki}}]{Ellis:1983ew}%
  \BibitemOpen
  \bibfield  {author} {\bibinfo {author} {\bibfnamefont {J.~R.}\ \bibnamefont
  {Ellis}}, \bibinfo {author} {\bibfnamefont {J.~S.}\ \bibnamefont {Hagelin}},
  \bibinfo {author} {\bibfnamefont {D.~V.}\ \bibnamefont {Nanopoulos}},
  \bibinfo {author} {\bibfnamefont {K.~A.}\ \bibnamefont {Olive}}, \ and\
  \bibinfo {author} {\bibfnamefont {M.}~\bibnamefont {Srednicki}},\ }\href
  {\doibase 10.1016/0550-3213(84)90461-9} {\bibfield  {journal} {\bibinfo
  {journal} {Nucl. Phys. B}\ }\textbf {\bibinfo {volume} {238}},\ \bibinfo
  {pages} {453} (\bibinfo {year} {1984})}\BibitemShut {NoStop}%
\bibitem [{\citenamefont {Scherrer}\ and\ \citenamefont
  {Turner}(1985)}]{Scherrer:1984fd}%
  \BibitemOpen
  \bibfield  {author} {\bibinfo {author} {\bibfnamefont {R.~J.}\ \bibnamefont
  {Scherrer}}\ and\ \bibinfo {author} {\bibfnamefont {M.~S.}\ \bibnamefont
  {Turner}},\ }\href {\doibase 10.1103/PhysRevD.31.681} {\bibfield  {journal}
  {\bibinfo  {journal} {Phys. Rev. D}\ }\textbf {\bibinfo {volume} {31}},\
  \bibinfo {pages} {681} (\bibinfo {year} {1985})}\BibitemShut {NoStop}%
\bibitem [{\citenamefont {Moroi}\ and\ \citenamefont
  {Randall}(2000)}]{Moroi:1999zb}%
  \BibitemOpen
  \bibfield  {author} {\bibinfo {author} {\bibfnamefont {T.}~\bibnamefont
  {Moroi}}\ and\ \bibinfo {author} {\bibfnamefont {L.}~\bibnamefont
  {Randall}},\ }\href {\doibase 10.1016/S0550-3213(99)00748-8} {\bibfield
  {journal} {\bibinfo  {journal} {Nucl. Phys. B}\ }\textbf {\bibinfo {volume}
  {570}},\ \bibinfo {pages} {455} (\bibinfo {year} {2000})},\ \Eprint
  {http://arxiv.org/abs/hep-ph/9906527} {arXiv:hep-ph/9906527} \BibitemShut
  {NoStop}%
\bibitem [{\citenamefont {Baltz}\ and\ \citenamefont
  {Murayama}(2003)}]{Baltz:2001rq}%
  \BibitemOpen
  \bibfield  {author} {\bibinfo {author} {\bibfnamefont {E.~A.}\ \bibnamefont
  {Baltz}}\ and\ \bibinfo {author} {\bibfnamefont {H.}~\bibnamefont
  {Murayama}},\ }\href {\doibase 10.1088/1126-6708/2003/05/067} {\bibfield
  {journal} {\bibinfo  {journal} {JHEP}\ }\textbf {\bibinfo {volume} {05}},\
  \bibinfo {pages} {067} (\bibinfo {year} {2003})},\ \Eprint
  {http://arxiv.org/abs/astro-ph/0108172} {arXiv:astro-ph/0108172} \BibitemShut
  {NoStop}%
\bibitem [{\citenamefont {Asaka}\ \emph {et~al.}(2006)\citenamefont {Asaka},
  \citenamefont {Shaposhnikov},\ and\ \citenamefont {Kusenko}}]{Asaka:2006ek}%
  \BibitemOpen
  \bibfield  {author} {\bibinfo {author} {\bibfnamefont {T.}~\bibnamefont
  {Asaka}}, \bibinfo {author} {\bibfnamefont {M.}~\bibnamefont {Shaposhnikov}},
  \ and\ \bibinfo {author} {\bibfnamefont {A.}~\bibnamefont {Kusenko}},\ }\href
  {\doibase 10.1016/j.physletb.2006.05.067} {\bibfield  {journal} {\bibinfo
  {journal} {Phys. Lett. B}\ }\textbf {\bibinfo {volume} {638}},\ \bibinfo
  {pages} {401} (\bibinfo {year} {2006})},\ \Eprint
  {http://arxiv.org/abs/hep-ph/0602150} {arXiv:hep-ph/0602150} \BibitemShut
  {NoStop}%
\bibitem [{\citenamefont {Bezrukov}\ \emph {et~al.}(2010)\citenamefont
  {Bezrukov}, \citenamefont {Hettmansperger},\ and\ \citenamefont
  {Lindner}}]{Bezrukov:2009th}%
  \BibitemOpen
  \bibfield  {author} {\bibinfo {author} {\bibfnamefont {F.}~\bibnamefont
  {Bezrukov}}, \bibinfo {author} {\bibfnamefont {H.}~\bibnamefont
  {Hettmansperger}}, \ and\ \bibinfo {author} {\bibfnamefont {M.}~\bibnamefont
  {Lindner}},\ }\href {\doibase 10.1103/PhysRevD.81.085032} {\bibfield
  {journal} {\bibinfo  {journal} {Phys. Rev. D}\ }\textbf {\bibinfo {volume}
  {81}},\ \bibinfo {pages} {085032} (\bibinfo {year} {2010})},\ \Eprint
  {http://arxiv.org/abs/0912.4415} {arXiv:0912.4415 [hep-ph]} \BibitemShut
  {NoStop}%
\bibitem [{\citenamefont {Nemev\v{s}ek}\ \emph {et~al.}(2012)\citenamefont
  {Nemev\v{s}ek}, \citenamefont {Senjanovi\'c},\ and\ \citenamefont
  {Zhang}}]{Nemevsek:2012cd}%
  \BibitemOpen
  \bibfield  {author} {\bibinfo {author} {\bibfnamefont {M.}~\bibnamefont
  {Nemev\v{s}ek}}, \bibinfo {author} {\bibfnamefont {G.}~\bibnamefont
  {Senjanovi\'c}}, \ and\ \bibinfo {author} {\bibfnamefont {Y.}~\bibnamefont
  {Zhang}},\ }\href {\doibase 10.1088/1475-7516/2012/07/006} {\bibfield
  {journal} {\bibinfo  {journal} {JCAP}\ }\textbf {\bibinfo {volume} {07}},\
  \bibinfo {pages} {006} (\bibinfo {year} {2012})},\ \Eprint
  {http://arxiv.org/abs/1205.0844} {arXiv:1205.0844 [hep-ph]} \BibitemShut
  {NoStop}%
\bibitem [{\citenamefont {Arcadi}\ and\ \citenamefont
  {Ullio}(2011)}]{Arcadi:2011ev}%
  \BibitemOpen
  \bibfield  {author} {\bibinfo {author} {\bibfnamefont {G.}~\bibnamefont
  {Arcadi}}\ and\ \bibinfo {author} {\bibfnamefont {P.}~\bibnamefont {Ullio}},\
  }\href {\doibase 10.1103/PhysRevD.84.043520} {\bibfield  {journal} {\bibinfo
  {journal} {Phys. Rev. D}\ }\textbf {\bibinfo {volume} {84}},\ \bibinfo
  {pages} {043520} (\bibinfo {year} {2011})},\ \Eprint
  {http://arxiv.org/abs/1104.3591} {arXiv:1104.3591 [hep-ph]} \BibitemShut
  {NoStop}%
\bibitem [{\citenamefont {Zhang}(2015)}]{Zhang:2015era}%
  \BibitemOpen
  \bibfield  {author} {\bibinfo {author} {\bibfnamefont {Y.}~\bibnamefont
  {Zhang}},\ }\href {\doibase 10.1088/1475-7516/2015/05/008} {\bibfield
  {journal} {\bibinfo  {journal} {JCAP}\ }\textbf {\bibinfo {volume} {05}},\
  \bibinfo {pages} {008} (\bibinfo {year} {2015})},\ \Eprint
  {http://arxiv.org/abs/1502.06983} {arXiv:1502.06983 [hep-ph]} \BibitemShut
  {NoStop}%
\bibitem [{\citenamefont {Patwardhan}\ \emph {et~al.}(2015)\citenamefont
  {Patwardhan}, \citenamefont {Fuller}, \citenamefont {Kishimoto},\ and\
  \citenamefont {Kusenko}}]{Patwardhan:2015kga}%
  \BibitemOpen
  \bibfield  {author} {\bibinfo {author} {\bibfnamefont {A.~V.}\ \bibnamefont
  {Patwardhan}}, \bibinfo {author} {\bibfnamefont {G.~M.}\ \bibnamefont
  {Fuller}}, \bibinfo {author} {\bibfnamefont {C.~T.}\ \bibnamefont
  {Kishimoto}}, \ and\ \bibinfo {author} {\bibfnamefont {A.}~\bibnamefont
  {Kusenko}},\ }\href {\doibase 10.1103/PhysRevD.92.103509} {\bibfield
  {journal} {\bibinfo  {journal} {Phys. Rev. D}\ }\textbf {\bibinfo {volume}
  {92}},\ \bibinfo {pages} {103509} (\bibinfo {year} {2015})},\ \Eprint
  {http://arxiv.org/abs/1507.01977} {arXiv:1507.01977 [astro-ph.CO]}
  \BibitemShut {NoStop}%
\bibitem [{\citenamefont {Soni}\ \emph {et~al.}(2017)\citenamefont {Soni},
  \citenamefont {Xiao},\ and\ \citenamefont {Zhang}}]{Soni:2017nlm}%
  \BibitemOpen
  \bibfield  {author} {\bibinfo {author} {\bibfnamefont {A.}~\bibnamefont
  {Soni}}, \bibinfo {author} {\bibfnamefont {H.}~\bibnamefont {Xiao}}, \ and\
  \bibinfo {author} {\bibfnamefont {Y.}~\bibnamefont {Zhang}},\ }\href
  {\doibase 10.1103/PhysRevD.96.083514} {\bibfield  {journal} {\bibinfo
  {journal} {Phys. Rev. D}\ }\textbf {\bibinfo {volume} {96}},\ \bibinfo
  {pages} {083514} (\bibinfo {year} {2017})},\ \Eprint
  {http://arxiv.org/abs/1704.02347} {arXiv:1704.02347 [hep-ph]} \BibitemShut
  {NoStop}%
\bibitem [{\citenamefont {Contino}\ \emph {et~al.}(2019)\citenamefont
  {Contino}, \citenamefont {Mitridate}, \citenamefont {Podo},\ and\
  \citenamefont {Redi}}]{Contino:2018crt}%
  \BibitemOpen
  \bibfield  {author} {\bibinfo {author} {\bibfnamefont {R.}~\bibnamefont
  {Contino}}, \bibinfo {author} {\bibfnamefont {A.}~\bibnamefont {Mitridate}},
  \bibinfo {author} {\bibfnamefont {A.}~\bibnamefont {Podo}}, \ and\ \bibinfo
  {author} {\bibfnamefont {M.}~\bibnamefont {Redi}},\ }\href {\doibase
  10.1007/JHEP02(2019)187} {\bibfield  {journal} {\bibinfo  {journal} {JHEP}\
  }\textbf {\bibinfo {volume} {02}},\ \bibinfo {pages} {187} (\bibinfo {year}
  {2019})},\ \Eprint {http://arxiv.org/abs/1811.06975} {arXiv:1811.06975
  [hep-ph]} \BibitemShut {NoStop}%
\bibitem [{\citenamefont {Evans}\ \emph {et~al.}(2020)\citenamefont {Evans},
  \citenamefont {Ghalsasi}, \citenamefont {Gori}, \citenamefont {Tammaro},\
  and\ \citenamefont {Zupan}}]{Evans:2019jcs}%
  \BibitemOpen
  \bibfield  {author} {\bibinfo {author} {\bibfnamefont {J.~A.}\ \bibnamefont
  {Evans}}, \bibinfo {author} {\bibfnamefont {A.}~\bibnamefont {Ghalsasi}},
  \bibinfo {author} {\bibfnamefont {S.}~\bibnamefont {Gori}}, \bibinfo {author}
  {\bibfnamefont {M.}~\bibnamefont {Tammaro}}, \ and\ \bibinfo {author}
  {\bibfnamefont {J.}~\bibnamefont {Zupan}},\ }\href {\doibase
  10.1007/JHEP02(2020)151} {\bibfield  {journal} {\bibinfo  {journal} {JHEP}\
  }\textbf {\bibinfo {volume} {02}},\ \bibinfo {pages} {151} (\bibinfo {year}
  {2020})},\ \Eprint {http://arxiv.org/abs/1910.06319} {arXiv:1910.06319
  [hep-ph]} \BibitemShut {NoStop}%
\bibitem [{\citenamefont {Cosme}\ \emph {et~al.}(2021)\citenamefont {Cosme},
  \citenamefont {Dutra}, \citenamefont {Ma}, \citenamefont {Wu},\ and\
  \citenamefont {Yang}}]{Cosme:2020mck}%
  \BibitemOpen
  \bibfield  {author} {\bibinfo {author} {\bibfnamefont {C.}~\bibnamefont
  {Cosme}}, \bibinfo {author} {\bibfnamefont {M.}~\bibnamefont {Dutra}},
  \bibinfo {author} {\bibfnamefont {T.}~\bibnamefont {Ma}}, \bibinfo {author}
  {\bibfnamefont {Y.}~\bibnamefont {Wu}}, \ and\ \bibinfo {author}
  {\bibfnamefont {L.}~\bibnamefont {Yang}},\ }\href {\doibase
  10.1007/JHEP03(2021)026} {\bibfield  {journal} {\bibinfo  {journal} {JHEP}\
  }\textbf {\bibinfo {volume} {03}},\ \bibinfo {pages} {026} (\bibinfo {year}
  {2021})},\ \Eprint {http://arxiv.org/abs/2003.01723} {arXiv:2003.01723
  [hep-ph]} \BibitemShut {NoStop}%
\bibitem [{\citenamefont {Dror}\ \emph {et~al.}(2020)\citenamefont {Dror},
  \citenamefont {Dunsky}, \citenamefont {Hall},\ and\ \citenamefont
  {Harigaya}}]{Dror:2020jzy}%
  \BibitemOpen
  \bibfield  {author} {\bibinfo {author} {\bibfnamefont {J.~A.}\ \bibnamefont
  {Dror}}, \bibinfo {author} {\bibfnamefont {D.}~\bibnamefont {Dunsky}},
  \bibinfo {author} {\bibfnamefont {L.~J.}\ \bibnamefont {Hall}}, \ and\
  \bibinfo {author} {\bibfnamefont {K.}~\bibnamefont {Harigaya}},\ }\href
  {\doibase 10.1007/JHEP07(2020)168} {\bibfield  {journal} {\bibinfo  {journal}
  {JHEP}\ }\textbf {\bibinfo {volume} {07}},\ \bibinfo {pages} {168} (\bibinfo
  {year} {2020})},\ \Eprint {http://arxiv.org/abs/2004.09511} {arXiv:2004.09511
  [hep-ph]} \BibitemShut {NoStop}%
\bibitem [{\citenamefont {Chanda}\ and\ \citenamefont
  {Unwin}(2021)}]{Chanda:2021tzi}%
  \BibitemOpen
  \bibfield  {author} {\bibinfo {author} {\bibfnamefont {P.}~\bibnamefont
  {Chanda}}\ and\ \bibinfo {author} {\bibfnamefont {J.}~\bibnamefont {Unwin}},\
  }\href {\doibase 10.1088/1475-7516/2021/06/032} {\bibfield  {journal}
  {\bibinfo  {journal} {JCAP}\ }\textbf {\bibinfo {volume} {06}},\ \bibinfo
  {pages} {032} (\bibinfo {year} {2021})},\ \Eprint
  {http://arxiv.org/abs/2102.02313} {arXiv:2102.02313 [hep-ph]} \BibitemShut
  {NoStop}%
\bibitem [{\citenamefont {Asadi}\ \emph {et~al.}(2021)\citenamefont {Asadi},
  \citenamefont {Slatyer},\ and\ \citenamefont {Smirnov}}]{Asadi:2021bxp}%
  \BibitemOpen
  \bibfield  {author} {\bibinfo {author} {\bibfnamefont {P.}~\bibnamefont
  {Asadi}}, \bibinfo {author} {\bibfnamefont {T.~R.}\ \bibnamefont {Slatyer}},
  \ and\ \bibinfo {author} {\bibfnamefont {J.}~\bibnamefont {Smirnov}},\
  }\href@noop {} {\bibfield  {journal} {\bibinfo  {journal} {arXiv}\ }
  (\bibinfo {year} {2021})},\ \Eprint {http://arxiv.org/abs/2111.11444}
  {arXiv:2111.11444 [hep-ph]} \BibitemShut {NoStop}%
\bibitem [{\citenamefont {Hasenkamp}\ and\ \citenamefont
  {Kersten}(2010)}]{Hasenkamp:2010if}%
  \BibitemOpen
  \bibfield  {author} {\bibinfo {author} {\bibfnamefont {J.}~\bibnamefont
  {Hasenkamp}}\ and\ \bibinfo {author} {\bibfnamefont {J.}~\bibnamefont
  {Kersten}},\ }\href {\doibase 10.1103/PhysRevD.82.115029} {\bibfield
  {journal} {\bibinfo  {journal} {Phys. Rev. D}\ }\textbf {\bibinfo {volume}
  {82}},\ \bibinfo {pages} {115029} (\bibinfo {year} {2010})},\ \Eprint
  {http://arxiv.org/abs/1008.1740} {arXiv:1008.1740 [hep-ph]} \BibitemShut
  {NoStop}%
\bibitem [{\citenamefont {Mohapatra}\ and\ \citenamefont
  {Pati}(1975)}]{Mohapatra:1974hk}%
  \BibitemOpen
  \bibfield  {author} {\bibinfo {author} {\bibfnamefont {R.~N.}\ \bibnamefont
  {Mohapatra}}\ and\ \bibinfo {author} {\bibfnamefont {J.~C.}\ \bibnamefont
  {Pati}},\ }\href {\doibase 10.1103/PhysRevD.11.566} {\bibfield  {journal}
  {\bibinfo  {journal} {Phys. Rev. D}\ }\textbf {\bibinfo {volume} {11}},\
  \bibinfo {pages} {566} (\bibinfo {year} {1975})}\BibitemShut {NoStop}%
\bibitem [{\citenamefont {Senjanovi\'c}(1979)}]{Senjanovic:1978ev}%
  \BibitemOpen
  \bibfield  {author} {\bibinfo {author} {\bibfnamefont {G.}~\bibnamefont
  {Senjanovi\'c}},\ }\href {\doibase 10.1016/0550-3213(79)90604-7} {\bibfield
  {journal} {\bibinfo  {journal} {Nucl. Phys. B}\ }\textbf {\bibinfo {volume}
  {153}},\ \bibinfo {pages} {334} (\bibinfo {year} {1979})}\BibitemShut
  {NoStop}%
\bibitem [{\citenamefont {Mohapatra}\ and\ \citenamefont
  {Senjanovi\'c}(1980)}]{Mohapatra:1979ia}%
  \BibitemOpen
  \bibfield  {author} {\bibinfo {author} {\bibfnamefont {R.~N.}\ \bibnamefont
  {Mohapatra}}\ and\ \bibinfo {author} {\bibfnamefont {G.}~\bibnamefont
  {Senjanovi\'c}},\ }\href {\doibase 10.1103/PhysRevLett.44.912} {\bibfield
  {journal} {\bibinfo  {journal} {Phys. Rev. Lett.}\ }\textbf {\bibinfo
  {volume} {44}},\ \bibinfo {pages} {912} (\bibinfo {year} {1980})}\BibitemShut
  {NoStop}%
\bibitem [{\citenamefont {Acharya}\ \emph {et~al.}(2009)\citenamefont
  {Acharya}, \citenamefont {Kane}, \citenamefont {Watson},\ and\ \citenamefont
  {Kumar}}]{Acharya:2009zt}%
  \BibitemOpen
  \bibfield  {author} {\bibinfo {author} {\bibfnamefont {B.~S.}\ \bibnamefont
  {Acharya}}, \bibinfo {author} {\bibfnamefont {G.}~\bibnamefont {Kane}},
  \bibinfo {author} {\bibfnamefont {S.}~\bibnamefont {Watson}}, \ and\ \bibinfo
  {author} {\bibfnamefont {P.}~\bibnamefont {Kumar}},\ }\href {\doibase
  10.1103/PhysRevD.80.083529} {\bibfield  {journal} {\bibinfo  {journal} {Phys.
  Rev. D}\ }\textbf {\bibinfo {volume} {80}},\ \bibinfo {pages} {083529}
  (\bibinfo {year} {2009})},\ \Eprint {http://arxiv.org/abs/0908.2430}
  {arXiv:0908.2430 [astro-ph.CO]} \BibitemShut {NoStop}%
\bibitem [{\citenamefont {Lesgourgues}(2011)}]{Lesgourgues:2011re}%
  \BibitemOpen
  \bibfield  {author} {\bibinfo {author} {\bibfnamefont {J.}~\bibnamefont
  {Lesgourgues}},\ }\href@noop {} {\bibfield  {journal} {\bibinfo  {journal}
  {arXiv}\ } (\bibinfo {year} {2011})},\ \Eprint
  {http://arxiv.org/abs/1104.2932} {arXiv:1104.2932 [astro-ph.IM]} \BibitemShut
  {NoStop}%
\bibitem [{\citenamefont {Blas}\ \emph {et~al.}(2011)\citenamefont {Blas},
  \citenamefont {Lesgourgues},\ and\ \citenamefont {Tram}}]{Blas:2011rf}%
  \BibitemOpen
  \bibfield  {author} {\bibinfo {author} {\bibfnamefont {D.}~\bibnamefont
  {Blas}}, \bibinfo {author} {\bibfnamefont {J.}~\bibnamefont {Lesgourgues}}, \
  and\ \bibinfo {author} {\bibfnamefont {T.}~\bibnamefont {Tram}},\ }\href
  {\doibase 10.1088/1475-7516/2011/07/034} {\bibfield  {journal} {\bibinfo
  {journal} {JCAP}\ }\textbf {\bibinfo {volume} {07}},\ \bibinfo {pages} {034}
  (\bibinfo {year} {2011})},\ \Eprint {http://arxiv.org/abs/1104.2933}
  {arXiv:1104.2933 [astro-ph.CO]} \BibitemShut {NoStop}%
\bibitem [{\citenamefont {Lesgourgues}\ and\ \citenamefont
  {Tram}(2011)}]{Lesgourgues:2011rh}%
  \BibitemOpen
  \bibfield  {author} {\bibinfo {author} {\bibfnamefont {J.}~\bibnamefont
  {Lesgourgues}}\ and\ \bibinfo {author} {\bibfnamefont {T.}~\bibnamefont
  {Tram}},\ }\href {\doibase 10.1088/1475-7516/2011/09/032} {\bibfield
  {journal} {\bibinfo  {journal} {JCAP}\ }\textbf {\bibinfo {volume} {09}},\
  \bibinfo {pages} {032} (\bibinfo {year} {2011})},\ \Eprint
  {http://arxiv.org/abs/1104.2935} {arXiv:1104.2935 [astro-ph.CO]} \BibitemShut
  {NoStop}%
\bibitem [{\citenamefont {Reid}\ \emph {et~al.}(2010)\citenamefont {Reid} \emph
  {et~al.}}]{Reid:2009xm}%
  \BibitemOpen
  \bibfield  {author} {\bibinfo {author} {\bibfnamefont {B.~A.}\ \bibnamefont
  {Reid}} \emph {et~al.},\ }\href {\doibase 10.1111/j.1365-2966.2010.16276.x}
  {\bibfield  {journal} {\bibinfo  {journal} {Mon. Not. Roy. Astron. Soc.}\
  }\textbf {\bibinfo {volume} {404}},\ \bibinfo {pages} {60} (\bibinfo {year}
  {2010})},\ \Eprint {http://arxiv.org/abs/0907.1659} {arXiv:0907.1659
  [astro-ph.CO]} \BibitemShut {NoStop}%
\bibitem [{\citenamefont {Lee}\ \emph {et~al.}(2013)\citenamefont {Lee} \emph
  {et~al.}}]{Lee:2012xb}%
  \BibitemOpen
  \bibfield  {author} {\bibinfo {author} {\bibfnamefont {K.-G.}\ \bibnamefont
  {Lee}} \emph {et~al.},\ }\href {\doibase 10.1088/0004-6256/145/3/69}
  {\bibfield  {journal} {\bibinfo  {journal} {Astron. J.}\ }\textbf {\bibinfo
  {volume} {145}},\ \bibinfo {pages} {69} (\bibinfo {year} {2013})},\ \Eprint
  {http://arxiv.org/abs/1211.5146} {arXiv:1211.5146 [astro-ph.CO]} \BibitemShut
  {NoStop}%
\bibitem [{\citenamefont {D'Eramo}\ and\ \citenamefont
  {Lenoci}(2021)}]{DEramo:2020gpr}%
  \BibitemOpen
  \bibfield  {author} {\bibinfo {author} {\bibfnamefont {F.}~\bibnamefont
  {D'Eramo}}\ and\ \bibinfo {author} {\bibfnamefont {A.}~\bibnamefont
  {Lenoci}},\ }\href {\doibase 10.1088/1475-7516/2021/10/045} {\bibfield
  {journal} {\bibinfo  {journal} {JCAP}\ }\textbf {\bibinfo {volume} {10}},\
  \bibinfo {pages} {045} (\bibinfo {year} {2021})},\ \Eprint
  {http://arxiv.org/abs/2012.01446} {arXiv:2012.01446 [hep-ph]} \BibitemShut
  {NoStop}%
\bibitem [{\citenamefont {Decant}\ \emph {et~al.}(2022)\citenamefont {Decant},
  \citenamefont {Heisig}, \citenamefont {Hooper},\ and\ \citenamefont
  {Lopez-Honorez}}]{Decant:2021mhj}%
  \BibitemOpen
  \bibfield  {author} {\bibinfo {author} {\bibfnamefont {Q.}~\bibnamefont
  {Decant}}, \bibinfo {author} {\bibfnamefont {J.}~\bibnamefont {Heisig}},
  \bibinfo {author} {\bibfnamefont {D.~C.}\ \bibnamefont {Hooper}}, \ and\
  \bibinfo {author} {\bibfnamefont {L.}~\bibnamefont {Lopez-Honorez}},\ }\href
  {\doibase 10.1088/1475-7516/2022/03/041} {\bibfield  {journal} {\bibinfo
  {journal} {JCAP}\ }\textbf {\bibinfo {volume} {03}},\ \bibinfo {pages} {041}
  (\bibinfo {year} {2022})},\ \Eprint {http://arxiv.org/abs/2111.09321}
  {arXiv:2111.09321 [astro-ph.CO]} \BibitemShut {NoStop}%
\bibitem [{\citenamefont {Mohapatra}\ and\ \citenamefont
  {Senjanovi\'c}(1981)}]{Mohapatra:1980yp}%
  \BibitemOpen
  \bibfield  {author} {\bibinfo {author} {\bibfnamefont {R.~N.}\ \bibnamefont
  {Mohapatra}}\ and\ \bibinfo {author} {\bibfnamefont {G.}~\bibnamefont
  {Senjanovi\'c}},\ }\href {\doibase 10.1103/PhysRevD.23.165} {\bibfield
  {journal} {\bibinfo  {journal} {Phys. Rev. D}\ }\textbf {\bibinfo {volume}
  {23}},\ \bibinfo {pages} {165} (\bibinfo {year} {1981})}\BibitemShut
  {NoStop}%
\bibitem [{\citenamefont {Nemev\v{s}ek}\ \emph {et~al.}(2013)\citenamefont
  {Nemev\v{s}ek}, \citenamefont {Senjanovi\'c},\ and\ \citenamefont
  {Tello}}]{Nemevsek:2012iq}%
  \BibitemOpen
  \bibfield  {author} {\bibinfo {author} {\bibfnamefont {M.}~\bibnamefont
  {Nemev\v{s}ek}}, \bibinfo {author} {\bibfnamefont {G.}~\bibnamefont
  {Senjanovi\'c}}, \ and\ \bibinfo {author} {\bibfnamefont {V.}~\bibnamefont
  {Tello}},\ }\href {\doibase 10.1103/PhysRevLett.110.151802} {\bibfield
  {journal} {\bibinfo  {journal} {Phys. Rev. Lett.}\ }\textbf {\bibinfo
  {volume} {110}},\ \bibinfo {pages} {151802} (\bibinfo {year} {2013})},\
  \Eprint {http://arxiv.org/abs/1211.2837} {arXiv:1211.2837 [hep-ph]}
  \BibitemShut {NoStop}%
\bibitem [{\citenamefont {Nemev\v{s}ek}\ and\ \citenamefont
  {Zhang}(2022)}]{wip2022}%
  \BibitemOpen
  \bibfield  {author} {\bibinfo {author} {\bibfnamefont {M.}~\bibnamefont
  {Nemev\v{s}ek}}\ and\ \bibinfo {author} {\bibfnamefont {Y.}~\bibnamefont
  {Zhang}},\ }\href@noop {} {\bibfield  {journal} {\bibinfo  {journal} {work in
  progress}\ } (\bibinfo {year} {2022})}\BibitemShut {NoStop}%
\bibitem [{\citenamefont {Keung}\ and\ \citenamefont
  {Senjanovi\'c}(1983)}]{Keung:1983uu}%
  \BibitemOpen
  \bibfield  {author} {\bibinfo {author} {\bibfnamefont {W.-Y.}\ \bibnamefont
  {Keung}}\ and\ \bibinfo {author} {\bibfnamefont {G.}~\bibnamefont
  {Senjanovi\'c}},\ }\href {\doibase 10.1103/PhysRevLett.50.1427} {\bibfield
  {journal} {\bibinfo  {journal} {Phys. Rev. Lett.}\ }\textbf {\bibinfo
  {volume} {50}},\ \bibinfo {pages} {1427} (\bibinfo {year}
  {1983})}\BibitemShut {NoStop}%
\bibitem [{\citenamefont {Nemev\v{s}ek}\ \emph {et~al.}(2011)\citenamefont
  {Nemev\v{s}ek}, \citenamefont {Nesti}, \citenamefont {Senjanovi\'c},\ and\
  \citenamefont {Zhang}}]{Nemevsek:2011hz}%
  \BibitemOpen
  \bibfield  {author} {\bibinfo {author} {\bibfnamefont {M.}~\bibnamefont
  {Nemev\v{s}ek}}, \bibinfo {author} {\bibfnamefont {F.}~\bibnamefont {Nesti}},
  \bibinfo {author} {\bibfnamefont {G.}~\bibnamefont {Senjanovi\'c}}, \ and\
  \bibinfo {author} {\bibfnamefont {Y.}~\bibnamefont {Zhang}},\ }\href
  {\doibase 10.1103/PhysRevD.83.115014} {\bibfield  {journal} {\bibinfo
  {journal} {Phys. Rev. D}\ }\textbf {\bibinfo {volume} {83}},\ \bibinfo
  {pages} {115014} (\bibinfo {year} {2011})},\ \Eprint
  {http://arxiv.org/abs/1103.1627} {arXiv:1103.1627 [hep-ph]} \BibitemShut
  {NoStop}%
\bibitem [{\citenamefont {Nemev\v{s}ek}\ \emph {et~al.}(2018)\citenamefont
  {Nemev\v{s}ek}, \citenamefont {Nesti},\ and\ \citenamefont
  {Popara}}]{Nemevsek:2018bbt}%
  \BibitemOpen
  \bibfield  {author} {\bibinfo {author} {\bibfnamefont {M.}~\bibnamefont
  {Nemev\v{s}ek}}, \bibinfo {author} {\bibfnamefont {F.}~\bibnamefont {Nesti}},
  \ and\ \bibinfo {author} {\bibfnamefont {G.}~\bibnamefont {Popara}},\ }\href
  {\doibase 10.1103/PhysRevD.97.115018} {\bibfield  {journal} {\bibinfo
  {journal} {Phys. Rev. D}\ }\textbf {\bibinfo {volume} {97}},\ \bibinfo
  {pages} {115018} (\bibinfo {year} {2018})},\ \Eprint
  {http://arxiv.org/abs/1801.05813} {arXiv:1801.05813 [hep-ph]} \BibitemShut
  {NoStop}%
\bibitem [{\citenamefont {Khachatryan}\ \emph {et~al.}(2017)\citenamefont
  {Khachatryan} \emph {et~al.}}]{CMS:2016ifc}%
  \BibitemOpen
  \bibfield  {author} {\bibinfo {author} {\bibfnamefont {V.}~\bibnamefont
  {Khachatryan}} \emph {et~al.} (\bibinfo {collaboration} {CMS}),\ }\href
  {\doibase 10.1016/j.physletb.2017.04.043} {\bibfield  {journal} {\bibinfo
  {journal} {Phys. Lett. B}\ }\textbf {\bibinfo {volume} {770}},\ \bibinfo
  {pages} {278} (\bibinfo {year} {2017})},\ \Eprint
  {http://arxiv.org/abs/1612.09274} {arXiv:1612.09274 [hep-ex]} \BibitemShut
  {NoStop}%
\bibitem [{\citenamefont {Aad}\ \emph {et~al.}(2019)\citenamefont {Aad} \emph
  {et~al.}}]{ATLAS:2019lsy}%
  \BibitemOpen
  \bibfield  {author} {\bibinfo {author} {\bibfnamefont {G.}~\bibnamefont
  {Aad}} \emph {et~al.} (\bibinfo {collaboration} {ATLAS}),\ }\href {\doibase
  10.1103/PhysRevD.100.052013} {\bibfield  {journal} {\bibinfo  {journal}
  {Phys. Rev. D}\ }\textbf {\bibinfo {volume} {100}},\ \bibinfo {pages}
  {052013} (\bibinfo {year} {2019})},\ \Eprint
  {http://arxiv.org/abs/1906.05609} {arXiv:1906.05609 [hep-ex]} \BibitemShut
  {NoStop}%
\bibitem [{\citenamefont {Aaboud}\ \emph {et~al.}(2019)\citenamefont {Aaboud}
  \emph {et~al.}}]{ATLAS:2019isd}%
  \BibitemOpen
  \bibfield  {author} {\bibinfo {author} {\bibfnamefont {M.}~\bibnamefont
  {Aaboud}} \emph {et~al.} (\bibinfo {collaboration} {ATLAS}),\ }\href
  {\doibase 10.1016/j.physletb.2019.134942} {\bibfield  {journal} {\bibinfo
  {journal} {Phys. Lett. B}\ }\textbf {\bibinfo {volume} {798}},\ \bibinfo
  {pages} {134942} (\bibinfo {year} {2019})},\ \Eprint
  {http://arxiv.org/abs/1904.12679} {arXiv:1904.12679 [hep-ex]} \BibitemShut
  {NoStop}%
\bibitem [{\citenamefont {Tumasyan}\ \emph {et~al.}(2022)\citenamefont
  {Tumasyan} \emph {et~al.}}]{CMS:2021dzb}%
  \BibitemOpen
  \bibfield  {author} {\bibinfo {author} {\bibfnamefont {A.}~\bibnamefont
  {Tumasyan}} \emph {et~al.} (\bibinfo {collaboration} {CMS}),\ }\href
  {\doibase 10.1007/JHEP04(2022)047} {\bibfield  {journal} {\bibinfo  {journal}
  {JHEP}\ }\textbf {\bibinfo {volume} {04}},\ \bibinfo {pages} {047} (\bibinfo
  {year} {2022})},\ \Eprint {http://arxiv.org/abs/2112.03949} {arXiv:2112.03949
  [hep-ex]} \BibitemShut {NoStop}%
\bibitem [{\citenamefont {Abdullahi}\ \emph {et~al.}(2022)\citenamefont
  {Abdullahi} \emph {et~al.}}]{Abdullahi:2022jlv}%
  \BibitemOpen
  \bibfield  {author} {\bibinfo {author} {\bibfnamefont {A.~M.}\ \bibnamefont
  {Abdullahi}} \emph {et~al.},\ }in\ \href@noop {} {\emph {\bibinfo {booktitle}
  {{2022 Snowmass Summer Study}}}}\ (\bibinfo {year} {2022})\ \Eprint
  {http://arxiv.org/abs/2203.08039} {arXiv:2203.08039 [hep-ph]} \BibitemShut
  {NoStop}%
\bibitem [{\citenamefont {Barbieri}\ and\ \citenamefont
  {Mohapatra}(1989)}]{Barbieri:1988av}%
  \BibitemOpen
  \bibfield  {author} {\bibinfo {author} {\bibfnamefont {R.}~\bibnamefont
  {Barbieri}}\ and\ \bibinfo {author} {\bibfnamefont {R.~N.}\ \bibnamefont
  {Mohapatra}},\ }\href {\doibase 10.1103/PhysRevD.39.1229} {\bibfield
  {journal} {\bibinfo  {journal} {Phys. Rev. D}\ }\textbf {\bibinfo {volume}
  {39}},\ \bibinfo {pages} {1229} (\bibinfo {year} {1989})}\BibitemShut
  {NoStop}%
\bibitem [{\citenamefont {Abazajian}\ \emph {et~al.}(2016)\citenamefont
  {Abazajian} \emph {et~al.}}]{CMB-S4:2016ple}%
  \BibitemOpen
  \bibfield  {author} {\bibinfo {author} {\bibfnamefont {K.~N.}\ \bibnamefont
  {Abazajian}} \emph {et~al.} (\bibinfo {collaboration} {CMB-S4}),\ }\href@noop
  {} {\bibfield  {journal} {\bibinfo  {journal} {arXiv}\ } (\bibinfo {year}
  {2016})},\ \Eprint {http://arxiv.org/abs/1610.02743} {arXiv:1610.02743
  [astro-ph.CO]} \BibitemShut {NoStop}%
\bibitem [{\citenamefont {Dodelson}\ and\ \citenamefont
  {Widrow}(1994)}]{Dodelson:1993je}%
  \BibitemOpen
  \bibfield  {author} {\bibinfo {author} {\bibfnamefont {S.}~\bibnamefont
  {Dodelson}}\ and\ \bibinfo {author} {\bibfnamefont {L.~M.}\ \bibnamefont
  {Widrow}},\ }\href {\doibase 10.1103/PhysRevLett.72.17} {\bibfield  {journal}
  {\bibinfo  {journal} {Phys. Rev. Lett.}\ }\textbf {\bibinfo {volume} {72}},\
  \bibinfo {pages} {17} (\bibinfo {year} {1994})},\ \Eprint
  {http://arxiv.org/abs/hep-ph/9303287} {arXiv:hep-ph/9303287} \BibitemShut
  {NoStop}%
\bibitem [{\citenamefont {De~Gouv\^ea}\ \emph {et~al.}(2020)\citenamefont
  {De~Gouv\^ea}, \citenamefont {Sen}, \citenamefont {Tangarife},\ and\
  \citenamefont {Zhang}}]{DeGouvea:2019wpf}%
  \BibitemOpen
  \bibfield  {author} {\bibinfo {author} {\bibfnamefont {A.}~\bibnamefont
  {De~Gouv\^ea}}, \bibinfo {author} {\bibfnamefont {M.}~\bibnamefont {Sen}},
  \bibinfo {author} {\bibfnamefont {W.}~\bibnamefont {Tangarife}}, \ and\
  \bibinfo {author} {\bibfnamefont {Y.}~\bibnamefont {Zhang}},\ }\href
  {\doibase 10.1103/PhysRevLett.124.081802} {\bibfield  {journal} {\bibinfo
  {journal} {Phys. Rev. Lett.}\ }\textbf {\bibinfo {volume} {124}},\ \bibinfo
  {pages} {081802} (\bibinfo {year} {2020})},\ \Eprint
  {http://arxiv.org/abs/1910.04901} {arXiv:1910.04901 [hep-ph]} \BibitemShut
  {NoStop}%
\bibitem [{\citenamefont {Kelly}\ \emph {et~al.}(2020)\citenamefont {Kelly},
  \citenamefont {Sen}, \citenamefont {Tangarife},\ and\ \citenamefont
  {Zhang}}]{Kelly:2020pcy}%
  \BibitemOpen
  \bibfield  {author} {\bibinfo {author} {\bibfnamefont {K.~J.}\ \bibnamefont
  {Kelly}}, \bibinfo {author} {\bibfnamefont {M.}~\bibnamefont {Sen}}, \bibinfo
  {author} {\bibfnamefont {W.}~\bibnamefont {Tangarife}}, \ and\ \bibinfo
  {author} {\bibfnamefont {Y.}~\bibnamefont {Zhang}},\ }\href {\doibase
  10.1103/PhysRevD.101.115031} {\bibfield  {journal} {\bibinfo  {journal}
  {Phys. Rev. D}\ }\textbf {\bibinfo {volume} {101}},\ \bibinfo {pages}
  {115031} (\bibinfo {year} {2020})},\ \Eprint
  {http://arxiv.org/abs/2005.03681} {arXiv:2005.03681 [hep-ph]} \BibitemShut
  {NoStop}%
\bibitem [{\citenamefont {Kelly}\ \emph {et~al.}(2021)\citenamefont {Kelly},
  \citenamefont {Sen},\ and\ \citenamefont {Zhang}}]{Kelly:2020aks}%
  \BibitemOpen
  \bibfield  {author} {\bibinfo {author} {\bibfnamefont {K.~J.}\ \bibnamefont
  {Kelly}}, \bibinfo {author} {\bibfnamefont {M.}~\bibnamefont {Sen}}, \ and\
  \bibinfo {author} {\bibfnamefont {Y.}~\bibnamefont {Zhang}},\ }\href
  {\doibase 10.1103/PhysRevLett.127.041101} {\bibfield  {journal} {\bibinfo
  {journal} {Phys. Rev. Lett.}\ }\textbf {\bibinfo {volume} {127}},\ \bibinfo
  {pages} {041101} (\bibinfo {year} {2021})},\ \Eprint
  {http://arxiv.org/abs/2011.02487} {arXiv:2011.02487 [hep-ph]} \BibitemShut
  {NoStop}%
\bibitem [{\citenamefont {Chichiri}\ \emph {et~al.}(2021)\citenamefont
  {Chichiri}, \citenamefont {Gelmini}, \citenamefont {Lu},\ and\ \citenamefont
  {Takhistov}}]{Chichiri:2021wvw}%
  \BibitemOpen
  \bibfield  {author} {\bibinfo {author} {\bibfnamefont {C.}~\bibnamefont
  {Chichiri}}, \bibinfo {author} {\bibfnamefont {G.~B.}\ \bibnamefont
  {Gelmini}}, \bibinfo {author} {\bibfnamefont {P.}~\bibnamefont {Lu}}, \ and\
  \bibinfo {author} {\bibfnamefont {V.}~\bibnamefont {Takhistov}},\ }\href@noop
  {} {\bibfield  {journal} {\bibinfo  {journal} {arXiv}\ } (\bibinfo {year}
  {2021})},\ \Eprint {http://arxiv.org/abs/2111.04087} {arXiv:2111.04087
  [hep-ph]} \BibitemShut {NoStop}%
\bibitem [{\citenamefont {Benso}\ \emph {et~al.}(2022)\citenamefont {Benso},
  \citenamefont {Rodejohann}, \citenamefont {Sen},\ and\ \citenamefont
  {Ramachandran}}]{Benso:2021hhh}%
  \BibitemOpen
  \bibfield  {author} {\bibinfo {author} {\bibfnamefont {C.}~\bibnamefont
  {Benso}}, \bibinfo {author} {\bibfnamefont {W.}~\bibnamefont {Rodejohann}},
  \bibinfo {author} {\bibfnamefont {M.}~\bibnamefont {Sen}}, \ and\ \bibinfo
  {author} {\bibfnamefont {A.~U.}\ \bibnamefont {Ramachandran}},\ }\href
  {\doibase 10.1103/PhysRevD.105.055016} {\bibfield  {journal} {\bibinfo
  {journal} {Phys. Rev. D}\ }\textbf {\bibinfo {volume} {105}},\ \bibinfo
  {pages} {055016} (\bibinfo {year} {2022})},\ \Eprint
  {http://arxiv.org/abs/2112.00758} {arXiv:2112.00758 [hep-ph]} \BibitemShut
  {NoStop}%
\bibitem [{\citenamefont {Ir\v{s}i\v{c}}\ \emph {et~al.}(2017)\citenamefont
  {Ir\v{s}i\v{c}} \emph {et~al.}}]{Irsic:2017ixq}%
  \BibitemOpen
  \bibfield  {author} {\bibinfo {author} {\bibfnamefont {V.}~\bibnamefont
  {Ir\v{s}i\v{c}}} \emph {et~al.},\ }\href {\doibase
  10.1103/PhysRevD.96.023522} {\bibfield  {journal} {\bibinfo  {journal} {Phys.
  Rev. D}\ }\textbf {\bibinfo {volume} {96}},\ \bibinfo {pages} {023522}
  (\bibinfo {year} {2017})},\ \Eprint {http://arxiv.org/abs/1702.01764}
  {arXiv:1702.01764 [astro-ph.CO]} \BibitemShut {NoStop}%
\bibitem [{\citenamefont {Nadler}\ \emph {et~al.}(2019)\citenamefont {Nadler},
  \citenamefont {Gluscevic}, \citenamefont {Boddy},\ and\ \citenamefont
  {Wechsler}}]{Nadler:2019zrb}%
  \BibitemOpen
  \bibfield  {author} {\bibinfo {author} {\bibfnamefont {E.~O.}\ \bibnamefont
  {Nadler}}, \bibinfo {author} {\bibfnamefont {V.}~\bibnamefont {Gluscevic}},
  \bibinfo {author} {\bibfnamefont {K.~K.}\ \bibnamefont {Boddy}}, \ and\
  \bibinfo {author} {\bibfnamefont {R.~H.}\ \bibnamefont {Wechsler}},\ }\href
  {\doibase 10.3847/2041-8213/ab1eb2} {\bibfield  {journal} {\bibinfo
  {journal} {Astrophys. J. Lett.}\ }\textbf {\bibinfo {volume} {878}},\
  \bibinfo {pages} {32} (\bibinfo {year} {2019})},\ \bibinfo {note} {[Erratum:
  Astrophys.J.Lett. 897, L46 (2020), Erratum: Astrophys.J. 897, L46 (2020)]},\
  \Eprint {http://arxiv.org/abs/1904.10000} {arXiv:1904.10000 [astro-ph.CO]}
  \BibitemShut {NoStop}%
\bibitem [{\citenamefont {Nadler}\ \emph {et~al.}(2021)\citenamefont {Nadler}
  \emph {et~al.}}]{DES:2020fxi}%
  \BibitemOpen
  \bibfield  {author} {\bibinfo {author} {\bibfnamefont {E.~O.}\ \bibnamefont
  {Nadler}} \emph {et~al.} (\bibinfo {collaboration} {DES}),\ }\href {\doibase
  10.1103/PhysRevLett.126.091101} {\bibfield  {journal} {\bibinfo  {journal}
  {Phys. Rev. Lett.}\ }\textbf {\bibinfo {volume} {126}},\ \bibinfo {pages}
  {091101} (\bibinfo {year} {2021})},\ \Eprint
  {http://arxiv.org/abs/2008.00022} {arXiv:2008.00022 [astro-ph.CO]}
  \BibitemShut {NoStop}%
\bibitem [{\citenamefont {Enzi}\ \emph {et~al.}(2021)\citenamefont {Enzi} \emph
  {et~al.}}]{Enzi:2020ieg}%
  \BibitemOpen
  \bibfield  {author} {\bibinfo {author} {\bibfnamefont {W.}~\bibnamefont
  {Enzi}} \emph {et~al.},\ }\href {\doibase 10.1093/mnras/stab1960} {\bibfield
  {journal} {\bibinfo  {journal} {Mon. Not. Roy. Astron. Soc.}\ }\textbf
  {\bibinfo {volume} {506}},\ \bibinfo {pages} {5848} (\bibinfo {year}
  {2021})},\ \Eprint {http://arxiv.org/abs/2010.13802} {arXiv:2010.13802
  [astro-ph.CO]} \BibitemShut {NoStop}%
\bibitem [{\citenamefont {Tremaine}\ and\ \citenamefont
  {Gunn}(1979)}]{TremaineGunn}%
  \BibitemOpen
  \bibfield  {author} {\bibinfo {author} {\bibfnamefont {S.}~\bibnamefont
  {Tremaine}}\ and\ \bibinfo {author} {\bibfnamefont {J.~E.}\ \bibnamefont
  {Gunn}},\ }\href {\doibase 10.1103/PhysRevLett.42.407} {\bibfield  {journal}
  {\bibinfo  {journal} {Phys. Rev. Lett.}\ }\textbf {\bibinfo {volume} {42}},\
  \bibinfo {pages} {407} (\bibinfo {year} {1979})}\BibitemShut {NoStop}%
\bibitem [{\citenamefont {Boyarsky}\ \emph {et~al.}(2009)\citenamefont
  {Boyarsky}, \citenamefont {Ruchayskiy},\ and\ \citenamefont
  {Iakubovskyi}}]{Boyarsky:2008ju}%
  \BibitemOpen
  \bibfield  {author} {\bibinfo {author} {\bibfnamefont {A.}~\bibnamefont
  {Boyarsky}}, \bibinfo {author} {\bibfnamefont {O.}~\bibnamefont
  {Ruchayskiy}}, \ and\ \bibinfo {author} {\bibfnamefont {D.}~\bibnamefont
  {Iakubovskyi}},\ }\href {\doibase 10.1088/1475-7516/2009/03/005} {\bibfield
  {journal} {\bibinfo  {journal} {JCAP}\ }\textbf {\bibinfo {volume} {03}},\
  \bibinfo {pages} {005} (\bibinfo {year} {2009})},\ \Eprint
  {http://arxiv.org/abs/0808.3902} {arXiv:0808.3902 [hep-ph]} \BibitemShut
  {NoStop}%
\bibitem [{\citenamefont {Domcke}\ and\ \citenamefont
  {Urbano}(2015)}]{Domcke:2014kla}%
  \BibitemOpen
  \bibfield  {author} {\bibinfo {author} {\bibfnamefont {V.}~\bibnamefont
  {Domcke}}\ and\ \bibinfo {author} {\bibfnamefont {A.}~\bibnamefont
  {Urbano}},\ }\href {\doibase 10.1088/1475-7516/2015/01/002} {\bibfield
  {journal} {\bibinfo  {journal} {JCAP}\ }\textbf {\bibinfo {volume} {01}},\
  \bibinfo {pages} {002} (\bibinfo {year} {2015})},\ \Eprint
  {http://arxiv.org/abs/1409.3167} {arXiv:1409.3167 [hep-ph]} \BibitemShut
  {NoStop}%
\bibitem [{\citenamefont {Di~Paolo}\ \emph {et~al.}(2018)\citenamefont
  {Di~Paolo}, \citenamefont {Nesti},\ and\ \citenamefont
  {Villante}}]{DiPaolo:2017geq}%
  \BibitemOpen
  \bibfield  {author} {\bibinfo {author} {\bibfnamefont {C.}~\bibnamefont
  {Di~Paolo}}, \bibinfo {author} {\bibfnamefont {F.}~\bibnamefont {Nesti}}, \
  and\ \bibinfo {author} {\bibfnamefont {F.~L.}\ \bibnamefont {Villante}},\
  }\href {\doibase 10.1093/mnras/sty091} {\bibfield  {journal} {\bibinfo
  {journal} {Mon. Not. Roy. Astron. Soc.}\ }\textbf {\bibinfo {volume} {475}},\
  \bibinfo {pages} {5385} (\bibinfo {year} {2018})},\ \Eprint
  {http://arxiv.org/abs/1704.06644} {arXiv:1704.06644 [astro-ph.GA]}
  \BibitemShut {NoStop}%
\bibitem [{\citenamefont {Ferraro}\ \emph {et~al.}(2022)\citenamefont
  {Ferraro}, \citenamefont {Sailer}, \citenamefont {Slosar},\ and\
  \citenamefont {White}}]{Ferraro:2022cmj}%
  \BibitemOpen
  \bibfield  {author} {\bibinfo {author} {\bibfnamefont {S.}~\bibnamefont
  {Ferraro}}, \bibinfo {author} {\bibfnamefont {N.}~\bibnamefont {Sailer}},
  \bibinfo {author} {\bibfnamefont {A.}~\bibnamefont {Slosar}}, \ and\ \bibinfo
  {author} {\bibfnamefont {M.}~\bibnamefont {White}},\ }\href@noop {}
  {\bibfield  {journal} {\bibinfo  {journal} {arXiv}\ } (\bibinfo {year}
  {2022})},\ \Eprint {http://arxiv.org/abs/2203.07506} {arXiv:2203.07506
  [astro-ph.CO]} \BibitemShut {NoStop}%
\bibitem [{\citenamefont {Chen}\ \emph {et~al.}(2016)\citenamefont {Chen},
  \citenamefont {Dvorkin}, \citenamefont {Huang}, \citenamefont {Namjoo},\ and\
  \citenamefont {Verde}}]{Chen:2016vvw}%
  \BibitemOpen
  \bibfield  {author} {\bibinfo {author} {\bibfnamefont {X.}~\bibnamefont
  {Chen}}, \bibinfo {author} {\bibfnamefont {C.}~\bibnamefont {Dvorkin}},
  \bibinfo {author} {\bibfnamefont {Z.}~\bibnamefont {Huang}}, \bibinfo
  {author} {\bibfnamefont {M.~H.}\ \bibnamefont {Namjoo}}, \ and\ \bibinfo
  {author} {\bibfnamefont {L.}~\bibnamefont {Verde}},\ }\href {\doibase
  10.1088/1475-7516/2016/11/014} {\bibfield  {journal} {\bibinfo  {journal}
  {JCAP}\ }\textbf {\bibinfo {volume} {11}},\ \bibinfo {pages} {014} (\bibinfo
  {year} {2016})},\ \Eprint {http://arxiv.org/abs/1605.09365} {arXiv:1605.09365
  [astro-ph.CO]} \BibitemShut {NoStop}%
\end{thebibliography}%

\newpage
\section{Supplemental Material}

In this supplement material, we provide more details on the derivation of several equations in the main text which are key to this analysis.
While in the main text those equations are presented at the order-of-magnitude level for clarity,
we derive more accurate prefactors here.

\subsection{A. Sudden decay approximation}

We give a brief overview of the sudden decay approximation, which is a good approximation and useful tool for exploring the late decay of long-lived particles in the early universe.
In the context of dark matter dilution mechanism discussed in this work, such an approximation allows us to derive and understand the 
important parametrical dependence in the final dark matter relic density $\Omega_X$, as well as $T_{\rm RH}$, the reheating temperature immediately after the dilutor decay.

Consider a dark matter particle $X$ (assuming it is a Majorana fermion) that freezes out from the Standard Model (SM) thermal plasma relativistically. Its yield $Y_X=n_X/s$, which is the ratio of its number density to the entropy density of the SM plasma, is given by
\begin{equation}\label{app1}
Y_X = \frac{135 \zeta(3)}{4\pi^4 g_*(T_{\rm fo})} \, .
\end{equation}
Here, $T_{\rm fo}$ is the photon temperature when dark matter freezes out, and $g_*(T_{\rm fo})$ counts the number of relativistic degrees of freedom in the SM plasma. Because most of our discussions will be restricted to temperatures above MeV scale (in order to ignite the big-bang nucleosynthesis successfully), we will not distinguish $g_*(T)$ and $g_{*S}(T)$ hereafter. %, unless necessary.
If nothing happens after freeze out, $Y_X$ would be a conserved quantity, and the dark matter relic density today would be 
\begin{equation}
\Omega^0_X = \frac{Y_X s_0 m_X}{\rho_0} \simeq 2.5 \left( \frac{m_X}{1\,\rm keV} \right) \left( \frac{100}{g_*(T_{\rm fo})} \right) \, ,
\end{equation}
where $s_0 = 2891.2\,{\rm cm}^{-3}$ is the entropy density in the universe today, and $\rho_0=1.05\times10^{-5} h^{2}\,{\rm GeV/cm^3}$ represents today's critical density with $h=0.674$.
In contrast, the observed value of $\Omega_X$ by {\it Planck} is 0.265. Because $m_X$ is constrained to be above several keV due to warm dark matter constraints,
the above result implies a dark matter overproduction problem.

To address the overproduction problem, we introduce a dilutor particle $Y$ (also assumed to be a fermion). It also freezes out relativistically and has a similar yield as Eq.~\eqref{app1} before decaying away. For sufficient entropy dilution, we further assume that $Y$ comes into matter domination of the total energy density of the universe before its decay, and most of its energy (entropy) is dumped to the SM sector. Under the sudden decay approximation, we have
\begin{equation}\label{app3}
\tau_Y^{-1} = H_{\rm before} = H_{\rm after} \, ,
\end{equation}
where $\tau_Y$ is the lifetime of $Y$, and $H_{\rm before,\, after}$ are the Hubble parameters ($H\equiv \sqrt{8\pi G\rho/3}$) immediately before and after the decay, respectively. In this approximation, the energy density of $Y$ immediately before its decay is equal to the radiation energy density immediately after.
The latter is given by $\rho_R = \pi^2 g_*(T_{\rm RH}) T_{\rm RH}^4/30$, where we introduced the ``reheating'' temperature $T_{\rm RH}$ of the SM plasma right after $Y$ decay.
The equation $\tau_Y^{-1} = H_{\rm after}$ leads to
\begin{equation}\label{app4}
\begin{split}
T_{\rm RH} &\simeq 0.78 \, g_*(T_{\rm RH})^{-1/4} \sqrt{\frac{M_{\rm pl}}{\tau_Y}} 
\\ &\simeq \frac{2.2 \,{\rm MeV}}{g_*(T_{\rm RH})^{1/4}} \sqrt{\frac{1\,\rm sec}{\tau_Y}} \, ,
\end{split}
\end{equation}
where $M_{\rm pl} = \sqrt{1/G_N} =1.2\times10^{19}\,$GeV is the Planck constant. 

In order to quantify the amount of dilution, we make use of the second equality in Eq.~\eqref{app3}, i.e. $\tau_Y^{-1}=H_{\rm before}$, where the energy density of $Y$ right before its decay is $\rho_Y = Y_Y s_{\rm before} m_Y$. Under our assumption of relativistic freeze-out of both $X$ and $Y$, we have $Y_Y =Y_X$ and $s_{\rm before}$ is the entropy density of the SM plasma before the decay occurs. Note however, that the dilutor $Y$ is behaving like matter at this moment. All of the above gives us
\begin{equation}\label{app5}
s_{\rm before} = \frac{\pi^3 g_*(T_{\rm fo})}{90 \zeta(3)} \frac{M_{\rm pl}^2}{m_Y \tau_Y^2} \, .
\end{equation}
On the other hand, the entropy density of the SM plasma immediately after $Y$ decay can be calculated using $T_{\rm RH}$ derived above,
\begin{equation}\label{app6}
s_{\rm after} = \frac{2\pi^2}{45} g_*(T_{\rm RH}) T_{\rm RH}^3 \, .
\end{equation}
With Eqs.~\eqref{app5} and \eqref{app6} we find the dilution factor $\mathcal{S}$ under the sudden decay approximation
\begin{equation}
\mathcal{S} \equiv \frac{s_{\rm after}}{s_{\rm before}} \simeq \frac{0.72 g_*(T_{\rm RH})^{1/4}}{g_*(T_{\rm fo})} \frac{m_Y \sqrt{\tau_Y}}{\sqrt{M_{\rm pl}}} \, .
\end{equation}

The diluted relic density of $X$ today is given by
\begin{equation}\label{app8}
%\begin{split}
\Omega_X = 
\frac{\Omega_X^0}{\mathcal{S}} 
%\simeq \frac{0.72 g_*(T_{\rm RH})^{1/4}}{g_*(T_{\rm fo})} \frac{m_Y \sqrt{\tau_Y}}{\sqrt{M_{\rm pl}}} 
%\\&
\simeq 0.26 \left( \frac{m_X}{1\,\rm keV} \right) \left(\frac{2.2\,\rm GeV}{m_Y}\right) \sqrt{\frac{1\,\rm sec}{\tau_Y}} \, .
%\end{split}
\end{equation}
This corresponds to Eq.~(2) in the main text, where we also included the possibility of $Y\to n X + m\, SM$ decay with a branching ratio ${\rm Br}_X$. Each decay of $Y$ can produce $n {\rm Br}_X$ secondary $X$ particles. Following the assumption of having equal number of $X$ and $Y$ to begin with (set by their relativistic freeze out condition), the $Y\to n X + m\, SM$ decay increases the net number of $X$ particles in the universe by a factor of $1+ n {\rm Br}_X$.

Requiring $\Omega_X$ to agree with the value measured by {\it Planck} allows to solve $\tau_Y$ in terms of $m_X$ and $m_Y$ using Eq.~\eqref{app8}. In turn, this allows us to rewrite Eq.~\eqref{app4} as
\begin{equation}\label{app9}
T_{\rm RH} \simeq \frac{1 \,{\rm MeV}}{g_*(T_{\rm RH})^{1/4}} \frac{m_Y}{10^6 m_X} \, .
\end{equation}
This corresponds to Eq.~(3) in the main text.

\subsection{B. Derivation of $T_{\rm NR}$}

Immediately after the $Y\to n X + m\, SM$ decay, each secondary $X$ particles roughly carries the energy $m_Y/(n+m)$. Under the sudden decay approximation, the corresponding temperature of the universe is given by $T_{\rm RH}$ in Eq.~\eqref{app4}. The energies of $X$ particles will redshift with the expansion of the universe. 
They start to turn non-relativistic when their energy drops below the mass.
This requires the scale factor of the universe to expand by a factor of
\begin{equation}
\frac{a_{\rm NR}}{a_{\rm RH}} = \frac{m_Y}{(n+m) \, m_X} \, .
\end{equation}
The corresponding temperature $T_{\rm NR}$ can be found with entropy conservation in the SM sector (note that $Y$ has already decayed away)
\begin{equation}
g_{*S}(T_{\rm NR}) T_{\rm NR}^3 a_{\rm NR}^3 = g_*(T_{\rm RH}) T_{\rm RH}^3 a_{\rm RH}^3 \, ,
\end{equation}
which leads to
\begin{equation}
\begin{split}
T_{\rm NR} &= T_{\rm RH} \left(\frac{g_*(T_{\rm RH})}{g_*(T_{\rm NR})} \right)^{1/3} \frac{a_{\rm RH}}{a_{\rm NR}} \\
& \simeq 0.63\,{\rm eV} \, (n+m) \, g_*(T_{\rm RH})^{\frac{1}{12}} \, .
\end{split}
\end{equation}
In the second step, we used Eq.~\eqref{app9} and $g_{*S}(T_{\rm NR})$ value $3.91$ valid for $T_{\rm NR}$ well below the electron mass. 
This corresponds to Eq.~(10) in the main text.

\subsection{C. Derivation of $\Delta N_{\rm eff}$}

Before the secondary $X$ particles from dilutor decay turn non-relativistic, they can contribute as additional radiation energy density of the universe.
This contribution effectively manifests as $\Delta N_{\rm eff}$, extra number of active neutrino species.

Immediately after dilutor $Y$ decay, the energy density ratio of secondary $X$ to the SM plasma is given by
\begin{equation}
\frac{\rho_X(T_{\rm RH})}{\rho_R(T_{\rm RH})} = \frac{y {\rm Br}_{X}}{1- y {\rm Br}_{X}} \, ,
\end{equation}
where $y$ is defined in Eq.~(5) and TABLE~I in the main text and denotes the the energy fraction carried by the $X$ particle(s) in each $Y\to nX(+m\,SM)$ decay.

To calculate the $\Delta N_{\rm eff}$, we need to find this ratio at a temperature around MeV, where the active neutrinos decouple from the thermal plasma. The energy density $\rho_X$ simply redshifts as $a^{-4}$, thus
\begin{equation}
\rho_X(T_{\rm MeV}) = \rho_X (T_{\rm RH}) \left( \frac{a_{\rm RH}}{a_{\rm MeV}} \right)^4 \, .
\end{equation}
For the SM plasma, energy is not conserved if $g_*$ varies with the temperature. Instead, entropy is conserved, which leads to
\begin{equation}
T_{\rm MeV} = T_{\rm RH} \left( \frac{a_{\rm RH}}{a_{\rm MeV}} \right) \left( \frac{g_*(T_{\rm RH})}{g_*(T_{\rm MeV})} \right)^{1/3} \, .
\end{equation}
The corresponding plasma radiation energy density is
\begin{equation}
\begin{split}
\rho_R(T_{\rm MeV}) &= \frac{\pi^2}{30} g_*(T_{\rm MeV}) T_{\rm RH}^4 \left( \frac{a_{\rm RH}}{a_{\rm MeV}} \right)^4 \left( \frac{g_*(T_{\rm RH})}{g_*(T_{\rm MeV})} \right)^{4/3} \\
&= \rho_R(T_{\rm RH}) \left( \frac{a_{\rm RH}}{a_{\rm MeV}} \right)^4 \left( \frac{g_*(T_{\rm RH})}{g_*(T_{\rm MeV})} \right)^{1/3} \, .
\end{split}
\end{equation}

At $T = \text{MeV}$, the energy fraction of a single flavor of active neutrino in the thermal plasma is
\begin{equation}
\frac{\rho_\nu(T_{\rm MeV})}{\rho_R(T_{\rm MeV})} = \frac{2\times \frac{7}{8}}{g_*(T_{\rm MeV})} = \frac{7}{43} \, ,
\end{equation}
where we used $g_*(T_{\rm MeV}) = 43/4$.

Therefore, the contribution of $X$ to $\Delta N_{\rm eff}$ is
\begin{equation}
\begin{split}
\Delta N_{\rm eff} &\equiv \frac{\rho_X(T_{\rm MeV})}{\rho_{\nu}(T_{\rm MeV})} = \frac{\rho_X(T_{\rm MeV})}{\rho_R(T_{\rm MeV})} \left(\frac{\rho_\nu(T_{\rm MeV})}{\rho_R(T_{\rm MeV})}\right)^{-1} \\
&= \frac{43}{7} \frac{y {\rm Br}_{X}}{1- y {\rm Br}_{X}} \left( \frac{g_*(T_{\rm MeV})}{g_*(T_{\rm RH})} \right)^{1/3} \\
&= \frac{43}{7} \frac{y {\rm Br}_{X}}{1- y {\rm Br}_{X}} \left( \frac{43}{4g_*(T_{\rm RH})} \right)^{1/3} \, .
\end{split}
\end{equation}
This corresponds to Eq.~(14) in the main text.

\subsection{D. A lower bound on $M_{W_R}$}

In the context of the LRSM, we consider the case where a right-handed neutrino $N_2$ both contributes to the type-I seesaw mechanism and serves as the dilutor $Y$.
The $\nu-N_2$ mixing is given by
\begin{equation}
\theta_{N_2\nu} \simeq \frac{m_\nu}{m_{N_2}} \simeq \frac{0.05\,\rm eV}{m_{N_2}} \, ,
\end{equation}
where we set the value $m_\nu$ to be the mass difference needed to explain the atmospheric neutrino oscillation.
Here we focus on the case $m_{N_2}<M_W$.
In this case, the three-body $N_2 \to \ell f \bar f'$ can take place via off-shell $W$-boson exchange, where $\ell$ is a charged lepton and $f, f'$ represent the fermion pair that couples to $W$.
This decay rate is
\begin{equation}
\Gamma_{N_2 \to \ell f \bar f'}= 9 \times \frac{\theta_{N_2\nu}^2 G_F^2 m_{N_2}^5}{192\pi^3} \, ,
\end{equation}
where the factor of 9 counts the final state multiplicity and we assumed $N_2$ is sufficiently heavy,
such that all the final fermion masses are neglected.

In contrast, the $N_2\to N_1\ell \ell$ decay occurs through the heavy $W_R$ boson exchange, 
where $N_1$ is the dark matter candidate. 
The decay rate is
\begin{equation}
  \Gamma_{N_2 \to N_1\ell \ell}= \frac{G_F^2 m_{N_2}^5}{192\pi^3} \left( \frac{M_W}{M_{W_R}} \right)^4 \, .
\end{equation}

The large scale structure constraint on dilutor to dark matter decay amounts to requiring
\begin{equation}
\Gamma_{N_2 \to N_1\ell \ell} \lesssim (1\%)\times \Gamma_{N_2 \to \ell f \bar f'} \, .
\end{equation}
This leads to
\begin{equation}
M_{W_R} > 55\,{\rm TeV} \left( \frac{m_{N_2}}{1\,{\rm GeV}} \right)^{1/4} \, ,
\end{equation}
which corresponds to Eq.~(13) in the main text.

\end{document}